\documentclass[journal=ancac3,manuscript=article]{achemso}
\usepackage[fontsize=11pt]{fontsize}
\usepackage[version=3]{mhchem} 
\usepackage{xcolor}

\author{Diogo V. Saraiva}
\author{Lotte Polling}
\author{Ivo R. Vermaire}
\author{Sander J. W. Vonk}
\author{Freddy T. Rabouw}
\author{Lisa Tran}
\email{l.tran@uu.nl}
\affiliation[Utrecht University]
{Soft Condensed Matter and Biophysics, Debye Institute for Nanomaterials Science, Utrecht University, 3584 CC Utrecht, The Netherlands}

\title[Title]
  {Controlling viscosity to engineer focal conic domains in photonic cellulose nanocrystal films}

\begin{document}

\begin{tocentry}
\includegraphics[width=1\linewidth]{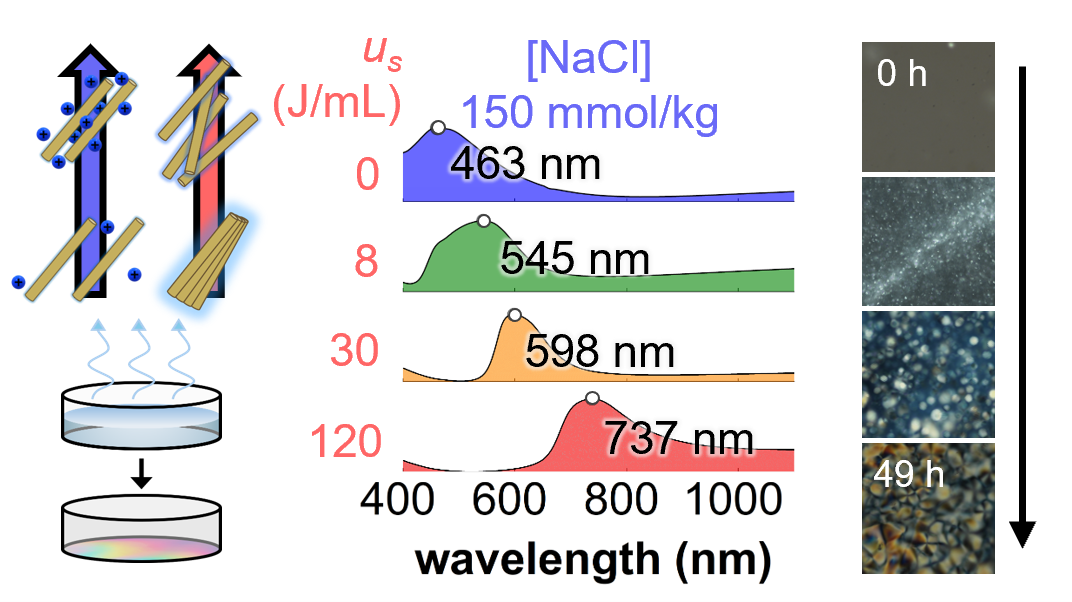}
\end{tocentry}

\begin{abstract}
Cellulose nanocrystals (CNCs) form cholesteric architectures that can have color specific reflectivity and enable sustainable photonic films. However, achieving uniform color, suppressing iridescence, and accessing ordered defect structures such as focal conic domains remain challenging. Here, we control the photonic properties of CNC films by steering the self assembly process. Across 24 dish-cast films with varying salt concentrations and sonication doses, we combine viscosity measurements, timelapse polarized optical microscopy, and angle-resolved reflectance spectroscopy to correlate evaporation dynamics with photonic structure. We show that viscosity, jointly controlled by \ce{NaCl}-mediated electrostatic screening and sonication-induced bundle fragmentation, dictates the extent of tactoid coalescence. Low-viscosity suspensions generate large, homogeneous cholesteric domains and narrow spectral responses, while high viscosity leads to arrested, heterogenous domains and increased diffuse light reflection. Critically, within a narrow parameter window of intermediate ionic strength and moderate sonication, we reproducibly engineer photonically active focal conic domains. These results identify viscosity-driven flow as a key, previously underappreciated factor in CNC self-assembly and establish design rules for producing structurally colored films with tunable photonic response, reduced iridescence, and controllable defect architectures.
\end{abstract}
\clearpage
\indent Structural color, originating from the interference between reflections of light on periodic or quasi-periodic nanostructures, has emerged as a promising mechanism for producing vibrant, pigment-free coloration. Its potential for use in sustainable coatings, optical sensors, and anti-counterfeiting technologies has stimulated extensive research interest \cite{saraiva_flexible_2024, chen_use_2016, yang_universal_2022}. Among various material systems, cellulose, an abundant and renewable biopolymer derived from plant biomass, offers a sustainable platform for the fabrication of structurally colored materials. The crystalline domains of cellulose fibrils can be isolated via acid hydrolysis to yield cellulose nanocrystals (CNCs), rod-like nanoparticles a few nanometers in width and hundreds of nanometers in length. When dispersed in water at appropriate concentrations, CNCs spontaneously self-assemble into a cholesteric liquid crystalline phase, whose helicoidal structure can be preserved upon solvent evaporation, producing iridescent solid films \cite{revol_helicoidal_1992}. 

The structural periodicity is quantified by the pitch ($p$), defined as the distance required for the rod-like nanocrystals to complete a $2\pi$ rotation (see Figure~\ref{ch2:fig1}A, right inset). The pitch determines the wavelength of selectively reflected light. From Bragg's law, in idealized, defect-free helical assemblies, the reflected wavelength ($\lambda$) is given by
\begin{equation}
    \lambda = np \sin(\theta)
    \label{eq:lambda}
\end{equation}
where $n$ is the average refractive index of the film ($\approx1.6$ for CNCs) \cite{parker_self-assembly_2018}, and $\theta$ is the angle of reflection relative to the film's surface \cite{de_vries_rotatory_1951}. Consequently, variations in $p$ or domain orientation typically give rise to strong angle-dependent color changes in cholesteric CNC films.

Evaporation-induced self-assembly of CNCs, typically performed in confined geometries such as petri dishes, proceeds through tactoid nucleation, coalescence, and eventual solidification, depicted schematically in Figure~\ref{ch2:fig1}A (right) \cite{tran_tactoid_2018}. The fully dried film consists of multiple cholesteric domains delineated by topological defects that originate during the coalescence of tactoids \cite{wang_structure_2016}. Each domain is characterized by a specific pitch, orientation, and lateral extent, collectively determining the film’s photonic response. Historically, optimization of CNC-based photonic materials has focused on promoting structural uniformity and minimizing defects to achieve narrow reflection bandwidths and high color saturation \cite{yao_flexible_2017}. Such uniformity has been pursued through strategies including reduction of CNC polydispersity \cite{honorato-rios_interrogating_2020}, modulation of solvent evaporation rate \cite{tran_tactoid_2018}, and application of shear to control helical alignment \cite{saha_photonic_2018, park_macroscopic_2014}.

While these approaches enhance structural order, they inherently preserve the angular dependence of the reflected color (Equation \ref{eq:lambda}). Iridescence (angle-dependent color) remains intrinsic to the cholesteric architecture, limiting the use of CNC films in applications that require a consistent appearance over wide viewing angles. Recent studies have highlighted that the deliberate introduction of controlled disorder in a photonic system can suppress angular color dependence \cite{ballato_tailoring_2000, michels-brito_bright_2022}. Controlled disorder, manifested through spatial variations in domain orientation, defect distribution, or pitch, can induce diffuse light scattering that averages out directional reflection, thereby producing noniridescent structural color. Such effects are widely observed in biological photonic systems, where partial disorder and hierarchical organization yield angle-independent coloration \cite{forster_biomimetic_2010, agez_multiwavelength_2017}. Consequently, the controlled manipulation of topological defects and domain heterogeneity in CNC assemblies presents a promising strategy to modulate photonic properties and achieve a uniform visual appearance.

Tuning the self-assembly pathway of CNCs through chemical and mechanical means enables control over the resulting structural hierarchy. The addition of electrolytes, such as sodium chloride (\ce{NaCl}), screens the electrostatic repulsion between CNCs, strengthening chiral interactions and decreasing the pitch size, $p$, depicted schematically in Figure~\ref{ch2:fig1}A (left)\cite{dong_effects_1996, dong_effect_1997, reid_effect_2017, honorato-rios_fractionation_2018}. Conversely, ultrasonication disrupts CNC bundles that act as chiral dopants, thereby weakening chiral interactions and increasing $p$ (Figure~\ref{ch2:fig1}A, center) \cite{beck_controlling_2011, parton_chiral_2022, gicquel_impact_2019}. Both modifications influence multiple stages of evaporation-driven self-assembly: from tactoid nucleation and coalescence to kinetic arrest, as well as the suspension’s rheological behavior \cite{gicquel_impact_2019, honorato-rios_fractionation_2018, parton_chiral_2022, beck_controlling_2011, dong_effects_1996, hirai_phase_2009, araki_effect_2001}. Although the effects of salt and sonication on CNC viscosity have been individually reported \cite{wu_estimation_2019, shafiei-sabet_rheology_2012, beuguel_ultrasonication_2018}, their combined influence and its implications for defect evolution and flow-induced ordering remain poorly understood.

In this work, we systematically investigate the interplay between ionic content and sonication in directing the self-assembly and final cholesteric structure of dish-cast CNC films. We produce an array of 24 CNC films under varying salt concentration and sonication dose and characterize their optical and structural properties using polarized optical microscopy, time-resolved imaging, and angle-resolved spectroscopy. Our observations reveal the emergence of convective flow cells during tactoid annealing, highlighting the pivotal role of solvent viscosity in defect evolution and domain organization. We demonstrate that optimizing salt concentration and sonication dose provides a means to control viscosity-driven flow dynamics and, consequently, the balance between structural order and controlled disorder. This study establishes a framework for engineering noniridescent structural color through the deliberate regulation of topological defects and domain heterogeneity in sustainable CNC-based photonic materials.

This work establishes a mechanistic framework linking physicochemical processing parameters, namely salt concentration, sonication dose, and solvent viscosity, to the evolution of defects, domain morphology, and optical response in cellulose nanocrystal (CNC) films. By revealing how flow-mediated self-assembly governs the emergence and stabilization of topological defects, we demonstrate that controlled disorder can be purposefully introduced to tune the resultant optical properties. Our results advance the understanding of how hierarchical organization in sustainable colloidal materials can be tuned through the subtle balance between order and disorder. Beyond elucidating the interplay between electrostatic screening, particle morphology, and hydrodynamic flow in colloidal self-assembly, this study provides design principles for tuning structural color through defect control, opening new avenues for scalable bio-derived photonic systems.

\clearpage

\section{Results and discussion}

\subsection{Tuning color with salt and sonication}
\begin{figure}[H]
\begin{center}
  \includegraphics[width=1.0\linewidth]{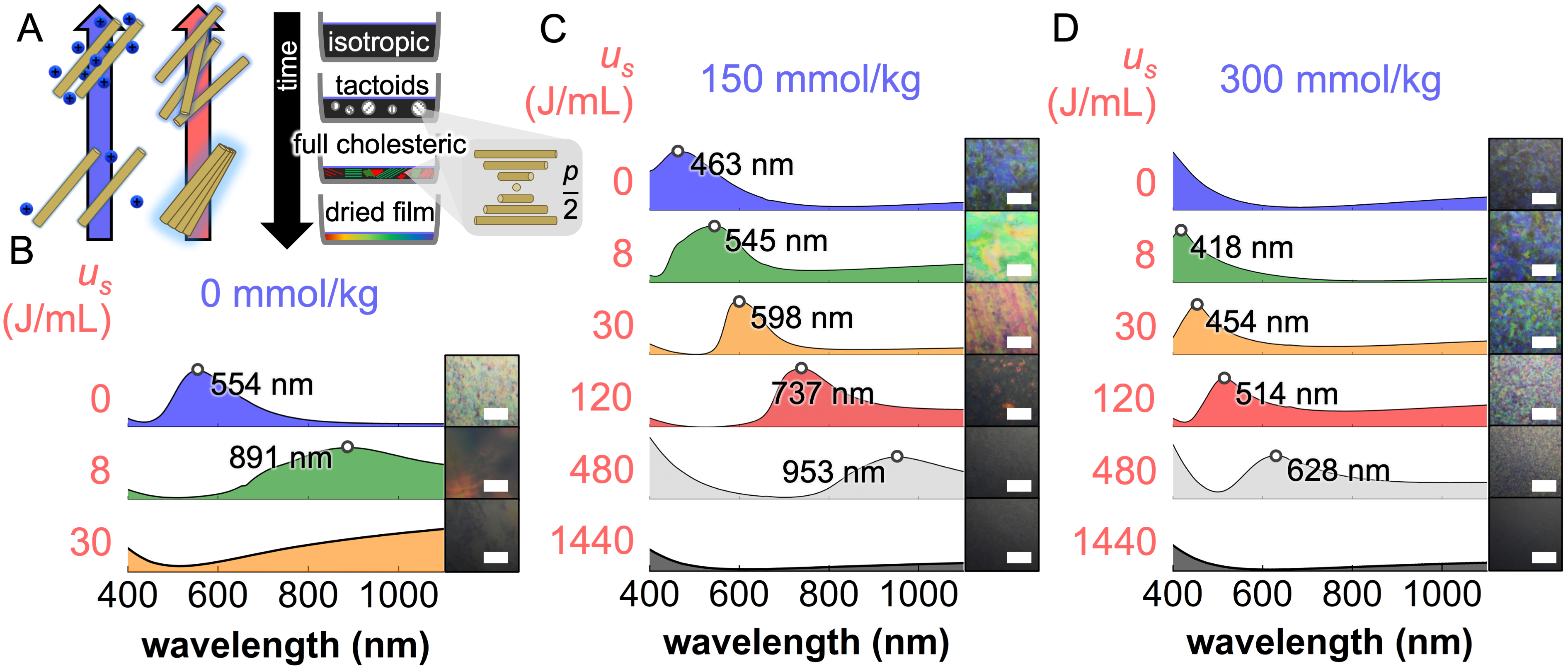}
  \end{center}
\caption{(A) Schematics illustrating how processing parameters influence CNC interactions and the resulting film color. Left: The blue arrow indicates that increasing salt concentration reduces the Debye length in solution, blue-shifting the reflected color of the dried film. Center: The red arrow indicates that increasing sonication dose fragments CNC bundles, red-shifting the reflected color. Right: Schematics of a CNC suspension evaporating in a petri dish with time progressing from top to bottom, leading to cholesteric self-assembly (with a half-pitch rotation of the CNCs depicted in the grey inset) and structural coloration in the final film. (B–D) Arrays of reflection-mode optical micrographs (captured with white light illumination through a left circular polarizer), each paired with a reflection spectrum from the same region. $\lambda_\text{max}$ is reported for each spectrum, denoted by a white dot. Panels are ordered by \ce{NaCl} concentration: (B) 0 mmol/kg of CNC, (C) 150 mmol/kg, and (D) 300 mmol/kg. Within each panel, spectra and images are arranged from top to bottom by sonication dose $u_s$ (0–1440 J/mL). Scale bar: 200 {\textmu}m.}\label{ch2:fig1}
\end{figure}

\indent \indent To investigate the effects of salt and sonication on the optical appearance of CNC films, we evaporated 24 CNC suspensions while tuning only their salt concentration, [\ce{NaCl}] (four values from 0 up to 450 mmol/kg of CNC), and their sonication dose, $u_s$ (six values from 0 up to 1440 J/mL). The CNCs used in this study are obtained from the acid hydrolysis of filter paper made from cotton, as detailed in the Methods section. The resulting CNC films exhibited photonic responses spanning from the UV to the IR. In Figure \ref{ch2:fig1}, we present micrographs of the films alongside their corresponding UV–Vis reflectance spectra, each with the peak wavelength $\lambda_\text{max}$ labeled. Reflection microscope and petri dish photographs of the films are provided in SI Figures \ref{si_lcp} and \ref{si_photo}.

As expected, increasing the sonication dose $u_s$ leads to an increase in $\lambda_\text{max}$, whereas increasing [\ce{NaCl}] results in a decrease in $\lambda_\text{max}$, as shown in the reflection spectra in Figure \ref{ch2:fig1}. In the absence of added salt, raising $u_s$ from 0 to 30 J/mL rapidly shifts $\lambda_\text{max}$ into the infrared (Figure \ref{ch2:fig1}B). With increasing salt concentration, from [\ce{NaCl}] $=$ 150 to 300 mmol/kg, the red-shift of $\lambda_{\text{max}}$ occurs more steadily, but an excessive sonication dose of 1440 J/mL eliminates any discernible $\lambda_\text{max}$, as shown at the bottom of Figures \ref{ch2:fig1}C and \ref{ch2:fig1}D.

Overall, the data in Figure \ref{ch2:fig1} indicate that increasing salt concentration diminishes the effectiveness of sonication in red-shifting the film's photonic response. At [\ce{NaCl}] $=$ 0 mmol/kg, the range of $\lambda_\text{max}$ spans more than 550 nm for 0 $\leq$ $u_s$ $\leq$ 30 J/mL (Figure \ref{ch2:fig1}B), while a salt concentration of [\ce{NaCl}] $=$ 150 mmol/kg results in an intermediate $\lambda_\text{max}$ range of about 430 nm for 0 $\leq$ $u_s$ $\leq$ 480 J/mL (Figure \ref{ch2:fig1}C). At [\ce{NaCl}] $=$ 300 mmol/kg, the range narrows to approximately 250 nm for 0 $\leq$ $u_s$ $\leq$ 480 J/mL (Figure \ref{ch2:fig1}D). Thus, sonication is most effective at red-shifting the photonic response under low-salt conditions and becomes progressively less effective as salt concentration increases.

This trend can be explained by the combined physical effects induced by salt and sonication. Increasing salt concentration shortens the Debye length and decreases the effective volume of each particle. Based on the pH and [NaCl] of CNC suspensions with varying concentrations up to 20 wt.\% (see SI Figure \ref{si_debye}), the Debye length in our evaporating CNC suspensions can range from 5.5 to 1 nm as salt concentration increases from 0 to 300 mmol/kg. A Debye length of 5.5 nm is significant relative to the typical CNC width of 5–25 nm measured by AFM (SI Figure \ref{si_afm}) \cite{parton_chiral_2022}. Meanwhile, ultrasonication breaks apart CNC bundles, increasing the total effective volume of the dispersed crystallites, including their ionic clouds. Once a bundle fragments, the combined effective volume of the individual CNCs exceeds that of the original aggregate. In low-salt media with a long Debye length, this increase in effective volume is amplified, decreasing packing efficiency and producing a red-shift in pitch. In contrast, when the Debye length is short due to high salt concentration, CNC particles can pack more efficiently into the cholesteric structure, leading to a blue-shift in pitch \cite{gicquel_impact_2019}.

\clearpage

\subsection{Effect on film morphology}
\begin{figure}[H]
\begin{center}
  \includegraphics[width=1\linewidth]{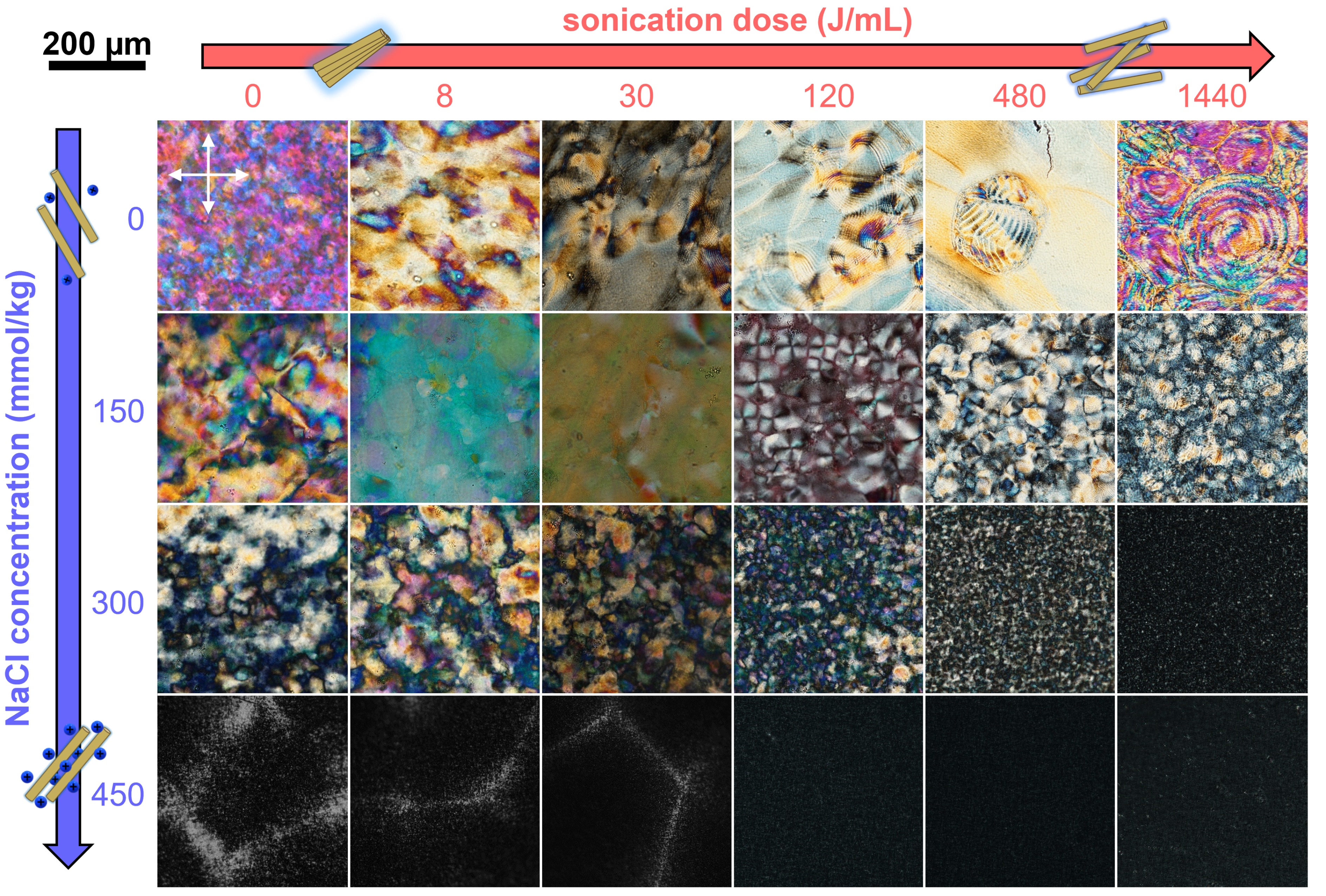}
  \end{center}
  \caption{An array of micrographs of CNC films taken in transmission mode, captured between crossed polarizers. Bright regions are birefringent and are thereby have anisotropic, liquid crystalline ordering, while dark regions are isotropic. The CNC films were made from suspensions prepared with varying amounts of \ce{NaCl} concentration (increasing from top to bottom) and sonication dose (increasing from left to right). Each image is taken at the center of its respective film (within a 2 mm radius).}
  \label{ch2:fig2}
\end{figure}

\indent \indent In addition to affecting the cholesteric pitch, salt and sonication also strongly influence the final film morphology. To examine these effects, we used polarized optical microscopy (POM) to observe films formed from solutions with varying sonication doses and salt concentrations. Figure \ref{ch2:fig2} displays all film morphologies in a 6$\times$4 grid of transmission images taken between crossed polarizers.

The cast films vary considerably in texture, particularly in domain size and density. In films with a salt concentration of 450 mmol/kg, regions that are dark under cross polarizers dominate, and only a few birefringent domains are retained, indicating that the film is largely in the isotropic phase by the end of the evaporation. Surprisingly, these small birefringent domains are organized into lines. For 300 and 450 mmol/kg, increasing sonication dose gradually reduces the size of the domains. At low salt concentrations (0 and 150 mmol/kg), more bright, birefringent regions are observed under crossed polarized microscopy, suggesting that (regardless of color) cholesteric structure is more readily obtained at these solution conditions, although the domain sizes and shapes vary widely. Notably, at 150 mmol/kg and sonication doses of 8 and 30 J/mL, the assembled films display minimal defects and large, homogeneous domains. Furthermore, Maltese crosses are observed at 150 mmol/kg and a sonication dose of 120 J/mL, which we attribute to the presence of focal conic domains (FCDs) -- regions of cholesteric ``layers" (defined by the half-pitch) that curve around focal defect lines \cite{yoon_three-dimensional_2013, tran_change_2017, roman_parabolic_2005}.

These observations reflect how salt concentration influences the onset of gelation and the development of cholesteric ordering. From previous studies, it is known that salt reduces the Debye length, lowering the CNC concentration at which kinetic arrest occurs while raising the critical CNC concentration for cholesteric nucleation \cite{honorato-rios_fractionation_2018, honorato-rios_equilibrium_2016}. As a result, the 450 mmol/kg suspension reaches kinetic arrest early, before a well-defined cholesteric phase can develop, leading to mostly dark isotropic regions with small amounts of birefringent domains. At 300 mmol/kg, kinetic arrest occurs slightly later, allowing more structured regions to form. At 150 and 0 mmol/kg, isotropic regions are less prominent because a larger fraction of the cholesteric structure is retained at the point of kinetic arrest.

When considering the effect of ultrasonication, films with [\ce{NaCl}] $=$ 300 and 450 mmol/kg show a decrease in average domain size with increasing sonication dose. This reduction may be associated with increases in the concentrations at which the cholesteric phase nucleates, forming birefringent tactoids, and at which the system is fully cholesteric as sonication increases \cite{gicquel_impact_2019}. In addition, the suspension may gel at earlier stages of tactoid formation \cite{parton_chiral_2022}, due to sonication fragmenting CNC particles and bundles.

At the higher salt concentrations, the decrease in domain size with increasing sonication dose $u_s$ suggests insufficient time for tactoid annealing. Sonication-induced fragmentation disrupts the formation of cholesteric order, while early gelation further restricts structural development. Consequently, higher sonication doses under these salt conditions lead to smaller cholesteric domains.

At the lower [\ce{NaCl}] values, the effect of sonication becomes more complex. Without added salt, increasing $u_s$ leads to larger birefringent regions of uniform color, indicating extended cholesteric regions with few domain walls. Fingerprint textures from the cholesteric helical structure become larger and more apparent with increasing $u_s$. At [\ce{NaCl}] $=$ 150 mmol/kg, films prepared with $u_s$ $=$ 8 and 30 J/mL are structurally homogeneous and show minimal defects. This homogeneity is supported by their particularly low spectral widths, obtained from the reflectance spectra in Figure \ref{ch2:fig1} (with values provided in SI Table \ref{si_fwhm}). Most strikingly, at $u_s$ $=$ 120 J/mL, Maltese crosses appear, indicating the formation of FCDs \cite{yoon_three-dimensional_2013, tran_change_2017, roman_parabolic_2005}. This represents the first reproducible observation of photonic FCDs in CNC films. Although FCDs have been previously reported in solid CNC films \cite{roman_parabolic_2005}, those systems exhibited large pitch sizes that prevented a photonic response and suffered from poor reproducibility. In contrast, here we observe reproducible FCD structures that reflect submicron light wavelengths ($\lambda_{\text{max}} = 737$ nm), but only within a narrow parameter space around [120 J/mL, 150 mmol/kg].

These results reveal that low salt levels allow sonication to promote, rather than inhibit, cholesteric ordering. Moderate sonication (8–30 J/mL) at 150 mmol/kg yields highly uniform films likely due to an optimal balance between tactoid formation/annealing and delayed gelation. At higher sonication (120 J/mL), the system accesses an alternative self-assembly pathway that produces FCDs.

\clearpage

\subsection{Focal conic domains}
\begin{figure}[H]
\begin{center}
\includegraphics[width=0.45\linewidth]{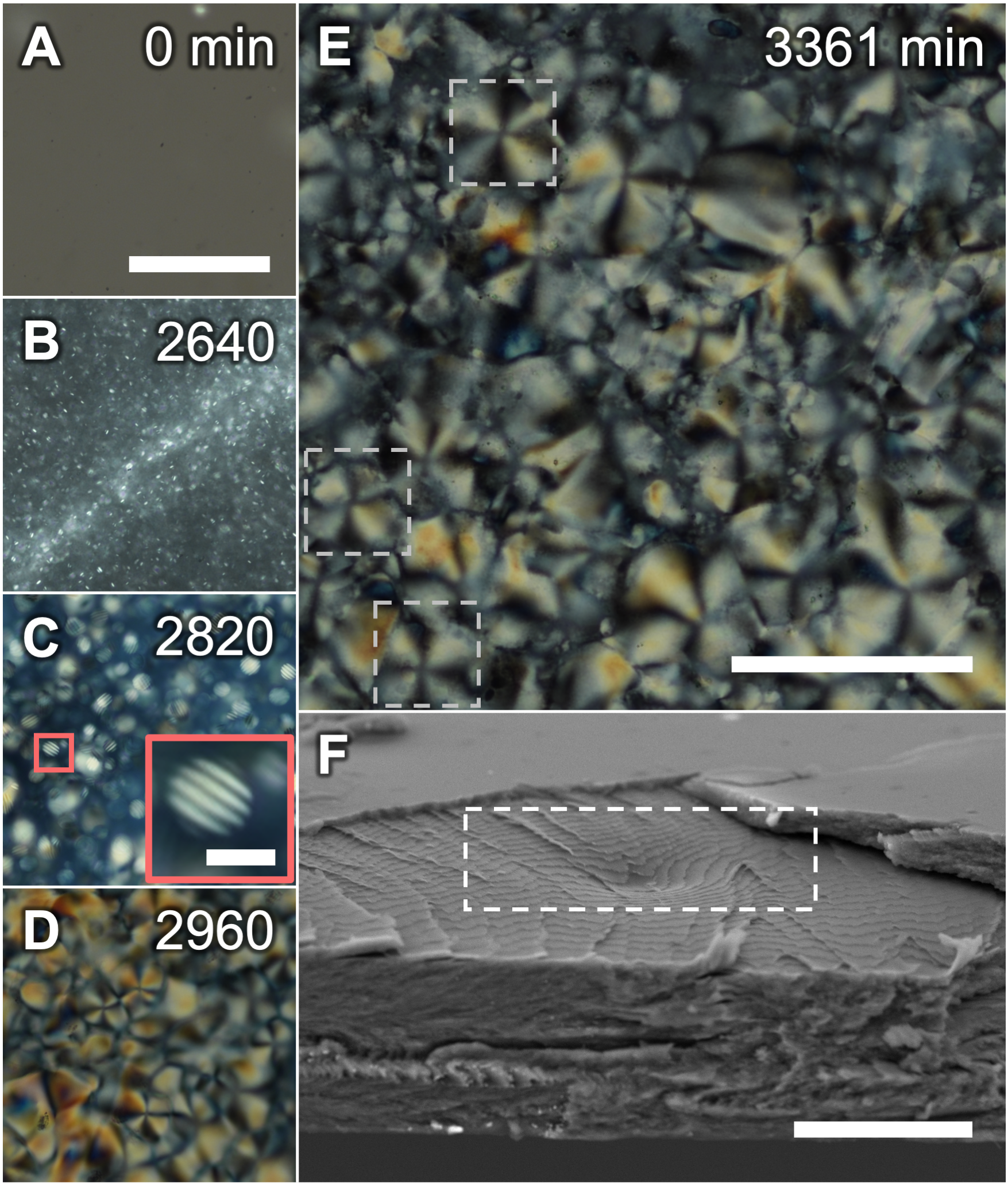}
\end{center}
\caption{(A–E) Frames from the timelapse video of a [120 J/mL, 150 mmol/kg] CNC suspension during evaporation. (A: 0 min) Evaporation begins with the solution in the isotropic regime; (B: 2640 min) the critical cholesteric concentration is reached, tactoids nucleate and organize into  lines; (C: 2820 min) tactoids grow and coalesce, a tactoid is highlighted in the solid red inset (scale bar: 10 {\textmu}m); (D: 2960 min) kinetic arrest is reached; (E: 3361 min) the film is compressed as evaporation is completed. FCDs are highlighted with gray dashed squares. (F) SEM image of the film cross-section and exposed glass–air interface. Scale bars: (A–E) 200 {\textmu}m, (C) 30 {\textmu}m.}
\label{ch2:fig3}
\end{figure}

\indent \indent To further investigate the self-assembly process leading to FCD formation under these optimized conditions ([120 J/mL, 150 mmol/kg]), we recorded a timelapse of film evaporation using POM. Figure \ref{ch2:fig3} presents selected frames from this timelapse (with the full sequence provided in SI Video 1), alongside cross-polarized optical micrographs and a scanning electron micrograph of a film cross-section.

Initially, the CNC suspension is isotropic (Figure \ref{ch2:fig3}A, 0 min). After several hours of gradual evaporation, tactoids begin to nucleate and align into linear arrangements (Figure \ref{ch2:fig3}B, 2640 min). As evaporation progresses, the cholesteric tactoids (identifiable from their characteristic fingerprint texture) grow and coalesce, shown in the micrograph of Figure \ref{ch2:fig3}C, 2820 min, with a tactoid highlighted in the inset. At around 2960 minutes, macroscopic flow visibly slows and eventually ceases. Note that this visual halt cannot be conclusively attributed to kinetic arrest; instead, it may reflect restricted mobility caused by a high volume fraction of cholesteric phase, which is known to have long relaxation times \cite{khadem_relaxation_2020}. Immediately following this event, FCDs emerge rapidly throughout the film. Afterward, only minimal structural changes are observed.

Interestingly, the evaporated films shown in Figure \ref{ch2:fig2} closely mirror the different stages observed in the timelapse. Early isotropic stages resemble films in the [NaCl] = 450 mmol/kg row with high sonication dose, where dark, isotropic regions are retained in the dried film. When tactoids first form and align in birefringent lines, the structure resembles the dried films in the 450 mmol/kg row with low sonication. In the 300 mmol/kg row, films from highest to lowest sonication resemble progressively later tactoid growth stages observed in the timelapse. Each stage of self-assembly, from isotropic to cholesteric, can be linked to a specific film in Figure \ref{ch2:fig2}, suggesting that salt and sonication systematically shift tactoid evolution and the point of kinetic arrest.

However, only one sample in Figure \ref{ch2:fig2}, [120 J/mL, 150 mmol/kg], retained FCDs. FCDs have been previously observed in CNC systems, both in capillaries \cite{elazzouzi_auto-organisation_2006} and in evaporated films, but with large pitch values (~3 microns) \cite{roman_parabolic_2005}. Although the mechanism of FCD formation was not fully explored in those works, it has been proposed \cite{frka-petesic_structural_2023} that compression-induced pitch contraction may trigger a Helfrich–Hurault (HH) instability \cite{hurault_static_1973, helfrich_deformation_1970, helfrich_electrohydrodynamic_1971, blanc_helfrich-hurault_2023}. In capillaries, FCDs appeared after one year of storage and further developed upon heating to 60$^\circ$C for one week. Both processes caused mild CNC desulfation, which reduced surface charge and contracted the cholesteric pitch \cite{frka-petesic_structural_2023}.

\begin{figure}[H]
\begin{center}
\includegraphics[width=0.92\linewidth]{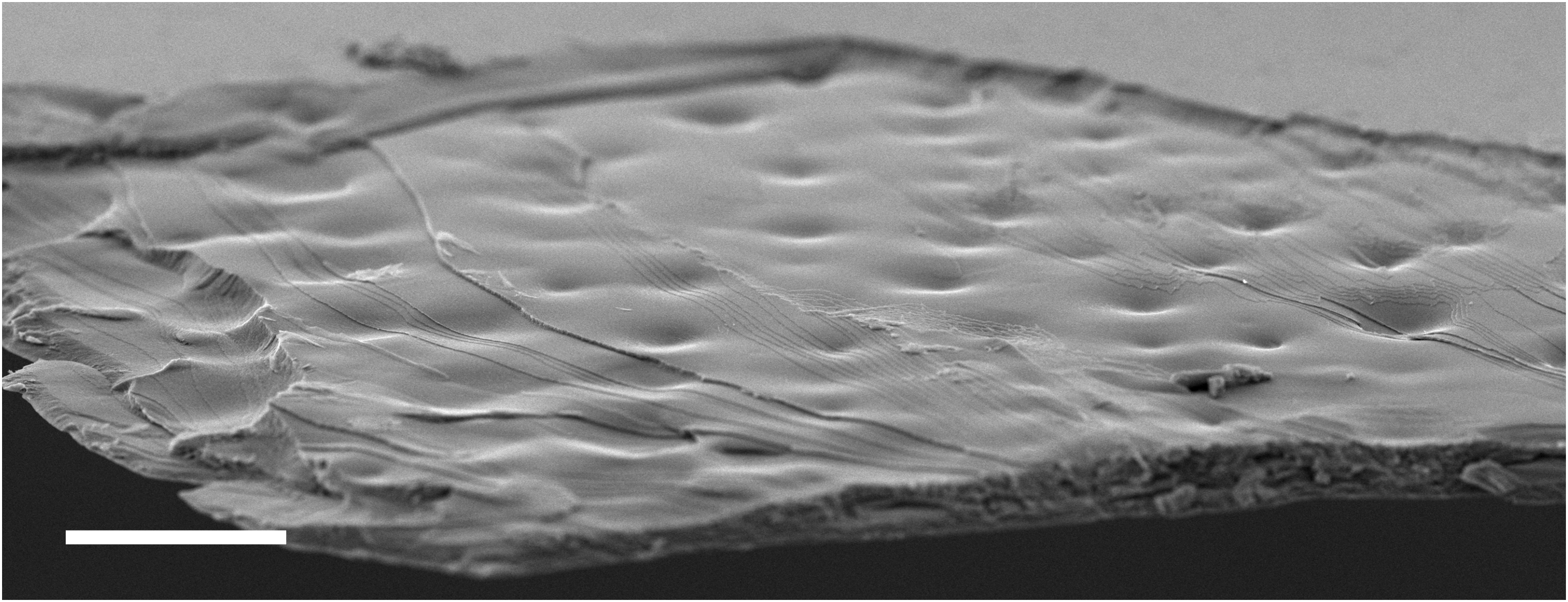}
\end{center}
\caption{SEM image of the cross-section and exposed glass–air interface of the [120 J/mL, 150 mmol/kg] film, showing multiple concavities indicative of FCDs. Scale bar: 100 {\textmu}m.}
\label{ch2:fig4}
\end{figure}

Compared to these previous reports, our system behaves distinctly: FCDs form only at a specific combination of salt concentration and sonication dose and have submicron pitch values that give rise to photonic properties. Supporting this, SEM imaging of the film cross-section (Figure \ref{ch2:fig3}F) reveals small undulations within the interior, consistent with focal conic structures. In another region, a network of concavities is observed beneath the surface (Figure \ref{ch2:fig4}). The diameter of these features (ranging from $\sim$10-100 microns) matches that of the FCDs observed using POM in Figure \ref{ch2:fig3}E. The film self-assembled between two flat interfaces (liquid–air and liquid–glass), both imposing planar anchoring. Despite this, internal undulations required for FCDs still formed within the bulk. These observations suggest that the FCDs are formed to relieve stresses within the film, similar to FCDs formed in molecular liquid crystal systems \cite{blanc_helfrich-hurault_2023, grritsma_electric-field-induced_1971}. 

We propose that the emergence of FCDs arises from a delayed kinetic arrest, which allows the cholesteric layers to remain deformable during the late stages of evaporation. When viscosity remains sufficiently low, the layers have not yet gelled and retain their liquid crystalline character, allowing for an elastic response to the continued evaporative stresses. These stresses force the deformable cholesteric layers to buckle and undulate, promoting the formation of FCDs. Only when the balance between ionic strength and sonication produces this delayed arrest do FCDs become energetically favorable: neither too early (which freezes the structure before undulation) nor too late (which prevents FCD formation by facilitating cholesteric annealing). Thus, we hypothesize that the balance between the self-assembly dynamics with the timing of kinetic arrest is crucial for enabling FCD formation in CNC films.

\clearpage

\subsection{Timing of self-assembly stages during evaporation}
\begin{figure}[H]
\begin{center}
  \includegraphics[width=1\linewidth]{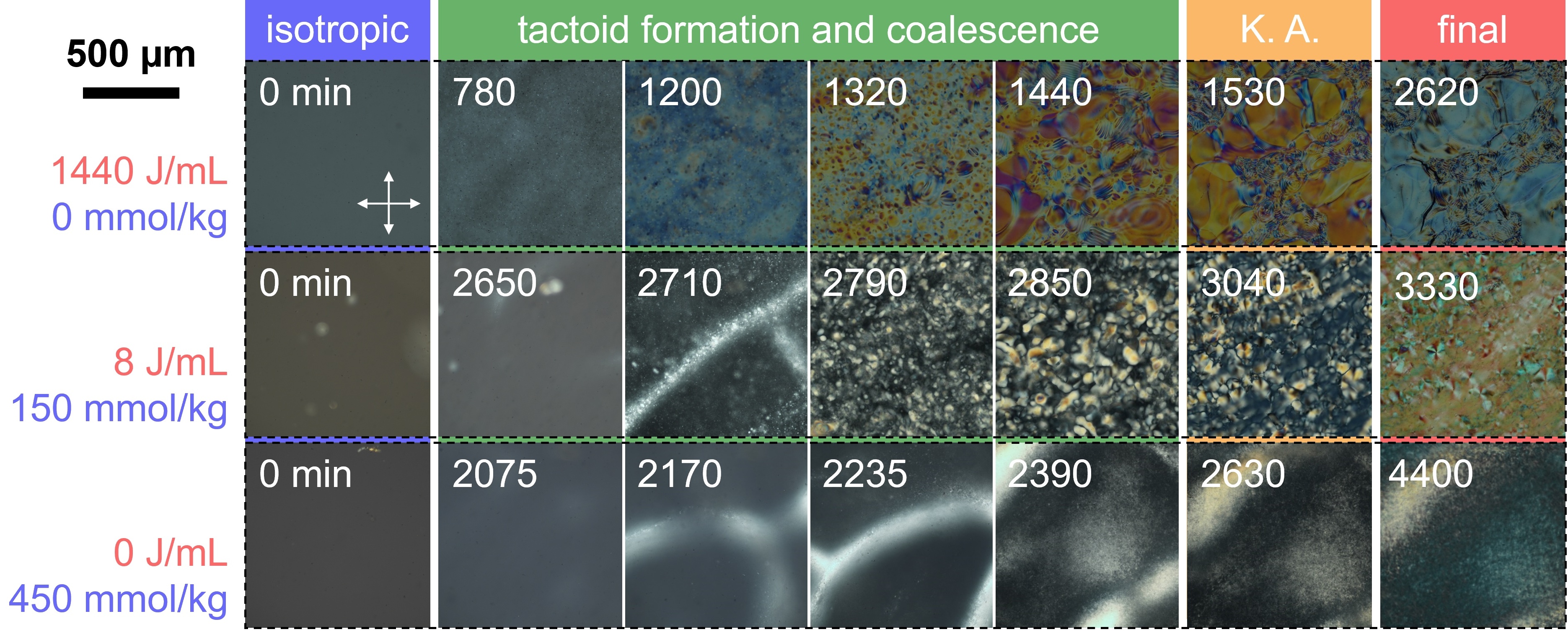}
  \end{center}
    \caption{Frames taken from three timelapse videos of evaporating CNC suspensions: [1440 J/mL, 0 mmol/kg], [8 J/mL, 150 mmol/kg], [0 J/mL, 450 mmol/kg]. All timelapse videos begin in the isotropic regime at [CNC] = 3 wt.\%. Tactoid formation and coalescence follow, and convection cells are observed in the bottom two samples. The second-to-last frame (K.A., kinetic arrest) marks the point at which tactoid motion ceases; however, this frame does not represent the exact moment of kinetic arrest, which cannot be determined visually as a lack of texture evolution can also arise from a high volume fraction of cholesteric phase. The final frame corresponds to complete evaporation. The scale bar for all frames is shown on the top left.}
  \label{ch2:fig5}
\end{figure}

\indent \indent To investigate how salt concentration and sonication dose modulate the progression of self-assembly stages and delay kinetic arrest, thereby promoting FCD formation, we examined the evaporation dynamics of selected samples using POM timelapse imaging. To better understand assembly dynamics during evaporation, we recorded timelapse videos of several CNC suspensions between crossed polarizers. Figure \ref{ch2:fig5} shows representative timelapse frames from three key samples that span the solution parameter space: [1440 J/mL, 0 mmol/kg], [8 J/mL, 150 mmol/kg], and [0 J/mL, 450 mmol/kg] (full videos in SI Videos 3, 4, and 5). As in Figure \ref{ch2:fig3}, all suspensions begin at [CNC] = 3 wt.\% in the isotropic phase. As evaporation proceeds, the concentration increases, and tactoids nucleate and drift across the field of view. For samples [8 J/mL, 150 mmol/kg] and [0 J/mL, 450 mmol/kg], the tactoids align into linear arrangements during nucleation. In contrast, tactoids in [1440 J/mL, 0 mmol/kg] form a homogeneous distribution without line patterns. With further CNC concentration increase, the line patterns in [8 J/mL, 150 mmol/kg] eventually disperse into a uniform tactoid distribution, whereas in [0 J/mL, 450 mmol/kg] the linear arrangements persist in the dried film.

To compare the stages of self-assembly, we examine the timestamps associated with key transitions. Sample [1440 J/mL, 0 mmol/kg] is the first to show tactoid nucleation (780 min), indicating that its critical cholesteric nucleation concentration is lowest. This is consistent with the long Debye length in salt-free suspensions, since effectively larger particles will interact at lower concentrations. Unexpectedly, [0 J/mL, 450 mmol/kg] is the second sample to nucleate the cholesteric phase (2075 min), which may be due to the presence of larger, unsonicated bundles promoting cholesteric phase nucleation. The sample [8 J/mL, 150 mmol/kg] nucleates the cholesteric phase the latest in the evaporation process. For the next stage of self-assembly, flow cessation occurs first in the [1440 J/mL, 0 mmol/kg] sample (1530 min). Sample [0 J/mL, 450 mmol/kg] ceases movement next, likely due to larger particle aggregates leading to gelation. Sample [8 J/mL, 150 mmol/kg] is the last to exhibit flow cessation. Evaporation completes in order of increasing salt concentration, which is consistent with the reduced evaporation rate of saline solutions \cite{mor_effect_2018}.

Overall, the samples exhibiting the longest duration between the nucleation of the first tactoid and flow cessation, from longest to shortest, are [1440 J/mL, 0 mmol/kg], [0 J/mL, 450 mmol/kg], and [8 J/mL, 150 mmol/kg]. However, the length of this relaxation period does not directly correspond to the size or morphology of the domains in the final film. This observation suggests that the specific nature of the dynamics and internal flow during this interval plays a critical role in determining the final structure.

To further understand the flow dynamics governing assembly, we examined the movement of tactoids in the timelapses. All samples exhibit a dominant radial flow directing CNC particles toward the Petri dish periphery. This capillary flow arises from higher evaporation rates at the perimeter relative to the center \cite{dumanli_digital_2014}. It produces the well-known coffee-ring effect, yielding a radial color gradient in freestanding films, as shown in SI Figure \ref{si_photo}. A second type of flow is observed when tactoids organize into lines. These line patterns are evident in Figure \ref{ch2:fig5} at 2710 min for [8 J/mL, 150 mmol/kg] and at 2170 min for [0 J/mL, 450 mmol/kg]. We hypothesize that this circulating flow originates from buoyancy-driven convection cells, where density differences between the isotropic and cholesteric phases cause tactoids to rise and sink. The tactoids accumulate at the cell boundaries, forming linear bands. Over time, if not inhibited by kinetic arrest, these line patterns dissolve into a more homogeneous cholesteric structure. Notably, convection cells were observed in all samples except those with no added salt.

To isolate the role of flow in tactoid assembly, we evaporated another [120 J/mL, 150 mmol/kg] suspension under identical conditions to Figure \ref{ch2:fig3}, but with the petri dish partially covered to slow down the evaporation. The full timelapse is provided in SI Figure \ref{si_slow}. Slowing evaporation from 2 to 11 days suppressed capillary and convective flows. Remarkably, the suspension self-assembled into numerous small tactoids that did not fully coalesce, producing a defect-rich structure. This result demonstrates that solvent flow is crucial in promoting tactoid coalescence and reducing defects during self-assembly.

Together, these observations shed light on why focal conic domains form only under specific conditions. Flow-driven tactoid coalescence is necessary to generate a coherent cholesteric structure. With sufficiently delayed kinetic arrest, as in the specific combination of 120 J/mL sonication and 150 mmol/kg \ce{NaCl}, the coherent cholesteric layers can exhibit a large scale elastic response to evaporative stresses. This process gives rise to the focal conic geometry observed exclusively under these conditions.

\clearpage

\subsection{Viscosity during self-assembly}

\begin{figure}[H]
\begin{center}
  \includegraphics[width=1\linewidth]{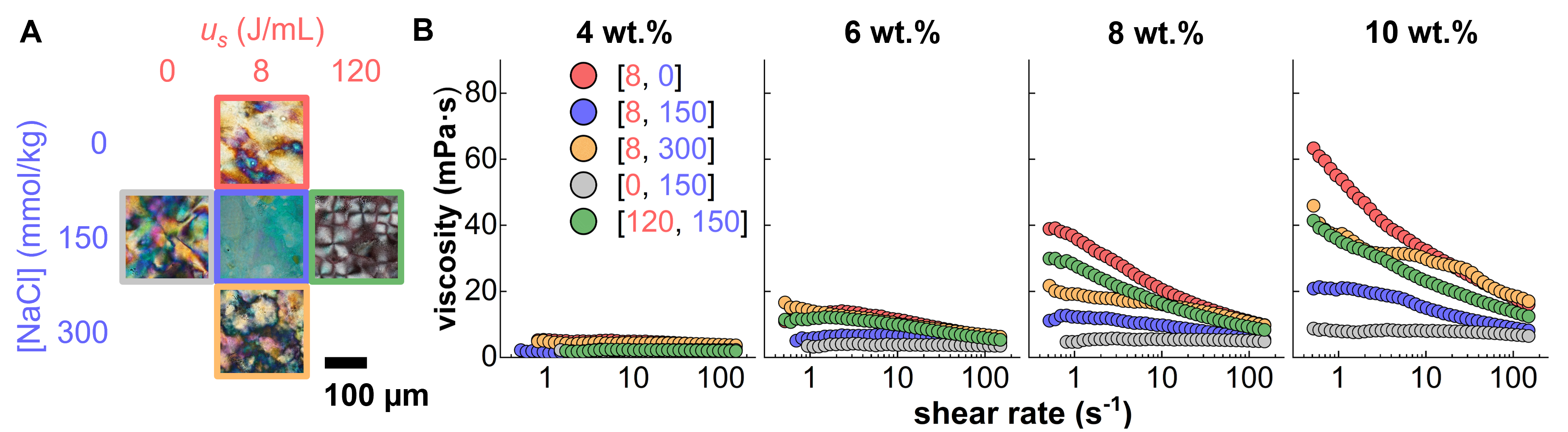}
  \end{center}
    \caption{(A) POM images of five CNC films, reproduced from Figure \ref{ch2:fig2} for ease of reference. (B) Flow curves of the selected CNC suspensions, measured at 4, 6, 8, and 10 wt.\% (left to right). Each curve in a given graph represents one of the five suspensions, labeled by the corresponding color used to outline the POM images in (A).}
  \label{ch2:fig6}
\end{figure}

\indent \indent To determine how viscosity governs the balance between flow, kinetic arrest, and the ability of cholesteric layers to deform into FCDs, we performed viscosity measurements on five suspensions with varying CNC concentrations, salt concentrations, and sonication doses, shown in Figure \ref{ch2:fig6}. Comparing samples with [\ce{NaCl}] $=$ 0, 150, and 300 mmol/kg at 8 and 10 wt.\% reveals that suspensions with 150 mmol/kg salt exhibit the lowest viscosity. Previous studies have shown that low \ce{NaCl} concentrations ($<$ 3 mM) reduce CNC viscosity, whereas high \ce{NaCl} concentrations ($>$ 5 mM) increases viscosity due to gradual aggregation \cite{wu_estimation_2019}. In our case, 150 mmol/kg corresponds to 4.5 mM at 3 wt.\% CNC, placing it within the reported viscosity minimum. The relatively low viscosity of 150 mmol/kg suspensions helps explain the uniform morphology observed in films at these conditions: a fluid medium facilitates tactoid fusion, suppressing defects. Conversely, the highest measured viscosity corresponds to [8 J/mL, 0 mmol/kg], the only sample that does not generate convection cells.

The flow curves also show that, at 150 mmol/kg [\ce{NaCl}], viscosity increases with sonication dose. While sonication is often reported to decrease viscosity by breaking aggregates \cite{shafiei-sabet_rheology_2012, beuguel_ultrasonication_2018}, those studies did not include electrolytes. In the presence of salt, higher sonication may increase particle number density, leading to mild electrostatic caging and higher viscosity. Atomic force microscopy and dynamic light scattering data (SI Figures \ref{si_afm} and \ref{si_dls}) confirm that particle size decreases with increasing sonication, as expected. This also explains why the [0 J/mL, 150 mmol/kg] film is structurally disordered despite its low viscosity: without sonication, large aggregates persist, hinder cholesteric annealing, and introduce defects. This interpretation is reinforced by comparing films made from filtered and unfiltered [0 J/mL, 150 mmol/kg] suspensions; the filtered sample exhibits slightly larger cholesteric domains (SI Figure \ref{si_filter}).

We fit the viscosity data of the 8 wt.-\% CNC suspensions in Figure \ref{ch2:fig6}B to the power-law model \cite{bird_dynamics_1987} describing the viscosity $\mu$ of non-Newtonian fluids:
\begin{equation}
\mu = K \left( \frac{\partial u}{\partial y} \right)^{n-1}
\label{eq:powerlaw}
\end{equation}
where $K$ is the flow consistency index, a quantity that reflects the average viscosity of the fluid, ${\partial u}/{\partial y}$ is the shear rate, and $n$ is the flow behavior index. The corresponding fits are shown in graphical form in Figure \ref{SI_viscosityfits}. At 8 wt.\%, all of the CNC suspensions have some fraction of the suspension in the cholesteric phase and exhibit shear-thinning behavior ($n < 1$), as expected for rod-like CNCs undergoing shear-induced alignment \cite{shafiei-sabet_rheology_2012, pal_rheology_2024}. Notably, [0~J/mL, 150~mmol/kg] has the highest $n$ value ($n = 0.985$), but also the lowest coefficient of determination $R^2$, indicating poor fit quality. This deviation likely stems from the absence of sonication.  While the power-law fluid model assumes a fluid with constant density, an unsonicated sample has large aggregated particles and thereby a high degree of heterogeneity.

\medskip \begin{table}[!ht]
    \centering  \begin{tabular}{|c|c|c|c|c|}
    \hline    \textbf{Sample} & $K$ (Pa$\cdot$s$^n$)&$n$&\textbf{$R^2$}& Ra\\ 
    \hline  [8 J/mL, 0 mmol/kg] & 0.038& 0.731&0.9946&218\\
    \hline    [8 J/mL, 150 mmol/kg] & 0.017& 0.831&0.9640& 728\\
    \hline    [8 J/mL, 300 mmol/kg] & 0.021& 0.872&0.9372& 416\\
    \hline    [0 J/mL, 150 mmol/kg] & 0.005& 0.985&0.2457& 1746\\
    \hline    [120 J/mL, 150 mmol/kg] & 0.028& 0.756&0.9995& 291\\
    \hline
    \end{tabular}
\caption{Power-law parameters ($K$, $n$, $R^2$) and estimated Rayleigh numbers (Ra) for the five CNC suspensions from Figure \ref{ch2:fig6} at 8 wt.\%.}
\label{ch2:viscosity_table}
\end{table}

To investigate the origin of convection cells, we estimated the solutal Rayleigh number (Ra), which quantifies the ratio of buoyant convection to diffusive transport. Higher Ra values indicate a greater likelihood of convection \cite{squires_microfluidics_2005}:
\begin{equation}
\text{Ra} = \frac{\Delta \rho l^3 g}{\mu D}
\label{eq:rayleigh}
\end{equation}
where $\Delta \rho$ is the density difference between isotropic and cholesteric phases (empirically measured in Supporting Information), $l^3$ is the suspension volume (3 mL), $\mu$ is the viscosity at 8 wt.\% (at a shear rate of $1$ s$^{-1}$), chosen to be a concentration that has both cholesteric and isotropic phases, $g$ is the gravitational acceleration, and $D$ is the diffusion coefficient of CNCs ($\approx$1 {\textmu}m$^2$/s) \cite{van_rie_anisotropic_2019}. The estimated Ra values are listed in Table \ref{ch2:viscosity_table}.

The suspension [8 J/mL, 0 mmol/kg] has the lowest Rayleigh number (218) and is the only sample that does not exhibit convection cells, consistent with its high viscosity. All other suspensions, which have higher Ra values, display convective flow during self-assembly (SI Videos 1, 3, 4, and 8). These observations indicate that the critical Rayleigh number for the onset of buoyancy-driven convection in CNC suspensions must lie between 218 and 291 (the value for [120 J/mL, 150 mmol/kg]). However, a precise prediction of this threshold is nontrivial: unlike classic Rayleigh–Bénard convection in simple Newtonian fluids, the instability here is triggered by a cholesteric phase transition and occurs in a highly anisotropic, non-Newtonian medium. Accurately modeling this behavior would require incorporating liquid-crystalline elasticity, anisotropic viscosity, and concentration-dependent density variations into the stability analysis. Furthermore, these convection cells are likely driven by buoyancy arising from local density differences during phase separation, rather than from thermal gradients, as in the classic model. Thus, viscosity governs tactoid coalescence by determining whether convective flows can develop during evaporative assembly.

However, viscosity also dictates when the suspension undergoes kinetic arrest: if viscosity rises too early (as in high-sonication or salt-free cases), the system gelates before coherent and large cholesteric domains are formed, preventing FCD formation. Conversely, if viscosity remains too low for too long (as in low-sonication cases), the system continues to anneal and defects are erased or prevented from forming rather than being trapped. Only at the intermediate viscosity achieved by the specific combination of 120 J/mL sonication dose and 150 mmol/kg [\ce{NaCl}] is kinetic arrest sufficiently delayed to allow deformable cholesteric layers to buckle into FCDs, explaining why FCDs form exclusively within this narrow parameter window.

\clearpage

\subsection{Structure and photonic response} 
\begin{figure}[H]
\begin{center}
\includegraphics[width=0.92\linewidth]{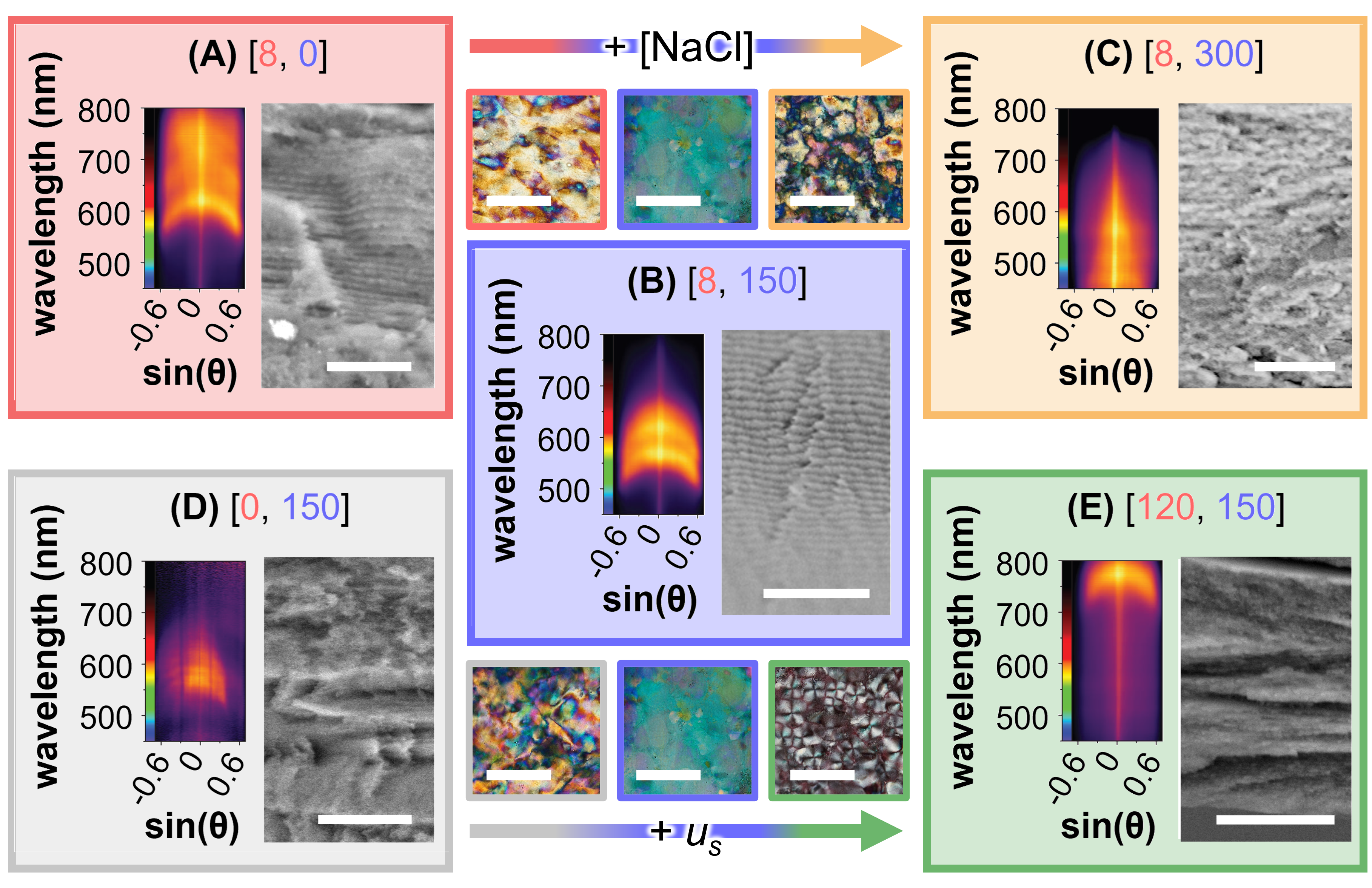}
\end{center}
\caption{(A–E) Angle-resolved reflectance spectra and scanning electron micrographs of five CNC films. For each film, the left side shows the angle-resolved reflectance spectrum collected from a 200~{\textmu}m region, with the left color bar denoting the corresponding color of the light wavelength and the right plot denoting the reflectivity ranging from 0 (black) to 100 \% (yellow). The right side shows an SEM image of an exposed film cross-section (scale bars: 10~{\textmu}m). The POM images from Figure \ref{ch2:fig2} at the top (corresponding to A, B, C) and bottom (corresponding to B, D, E) depict the film structure as seen through POM (Scale bars: 200~{\textmu}m), with color-coded outlines corresponding to panels A–E.}
\label{ch2:fig7}
\end{figure}

\indent \indent To further investigate how domain morphology influences the optical response, angle-resolved reflectance spectroscopy was performed on each film. Spectra were collected from a 200~{\textmu}m diameter region on the film surface (Figure \ref{ch2:fig7}, left side of each colored panel). To correlate optical behavior with internal structure, corresponding cross-sections were imaged using SEM (Figure \ref{ch2:fig7}, right side of each panel).

For each film, we examined both its photonic response and cholesteric domain morphology. In the angle-resolved spectra (Figure \ref{ch2:fig7}, left side of panels), the extent of more bright regions across wavelengths represent the spectral width, which indicates the range of reflected wavelengths. Additional spectra taken within both 20 and 200~{\textmu}m circular areas of the sample are provided in SI Figure \ref{SI_allangleresolved}. Films [8 J/mL, 0 mmol/kg] and [8 J/mL, 300 mmol/kg] (Figure \ref{ch2:fig7}A, C) display broader spectral widths, while [8 J/mL, 150 mmol/kg] and [120 J/mL, 150 mmol/kg] (Figure \ref{ch2:fig7}B, E) exhibit much narrower ranges. These trends correspond directly to the SEM observations (right panels): films with broader spectra show greater structural disorder and varying pitch values, whereas those with narrow spectra have cohesive, well-organized domains. Notably, [8 J/mL, 150 mmol/kg] shows an exceptionally uniform cross-section spanning several hundred microns.

Most spectra display a concave shape. This is expected, as planar cholesteric films exhibit angle-dependent coloration, as described by Equation \ref{eq:lambda}, and these films are cast on petri dishes that promote planar anchoring of the CNCs. Films with large domains and long-range order, such as [8 J/mL, 150 mmol/kg], have reflectivity maximums that steeply blueshift for larger incoming angles (Figure \ref{ch2:fig7}B). In contrast, films with small domains and significant disorder, such as [8 J/mL, 300 mmol/kg], have reflectivity maximums independent of the incoming angle (Figure \ref{ch2:fig7}C). Several spectra also exhibit higher-order reflection bands (Figure \ref{ch2:fig7}A, B, D), as previously reported in CNC films \cite{frka-petesic_angular_2019}.

Salt concentration has a strong influence on photonic response. Samples with either 0 or 300 mmol/kg \ce{NaCl} show wider spectral widths, while 150 mmol/kg yields a narrow spectrum. This agrees with our earlier findings that 150 mmol/kg promotes long-range cholesteric order with uniform pitch values. At this intermediate salt concentration, viscosity is minimized, allowing tactoids to coalesce into large domains that reflect a narrow band of wavelengths. In comparison, sonication dose plays a subtler role. Among the three films at 150 mmol/kg (Figure \ref{ch2:fig7}B, D, E), all display relatively narrow spectral widths, with only modest red-shifts at higher sonication. However, [0 J/mL, 150 mmol/kg] uniquely shows a mismatch between spectra taken at small and large scales, reflecting its broad distribution of domain sizes and colors observed in POM (Figure \ref{ch2:fig2}). To complement the spectra in Figure \ref{ch2:fig7}, we also quantified spectral widths from UV–Vis reflectance (SI Table \ref{si_fwhm}), which confirm minimal spectral widths for films at $u_s = 8$ and 30 J/mL.

Among all samples, the film containing FCDs ([120 J/mL, 150 mmol/kg], Figure \ref{ch2:fig7}E) exhibits a unique optical response. Despite the presence of FCDs, the spectral width of this film is the narrowest in Figure \ref{ch2:fig7}E, indicating a well-defined cholesteric pitch and strong structural coherence across the domains. A cholesteric with a well-defined pitch value would more readily resist layer deformation from applied stresses, yielding the Helfrich-Hurault instability, in which deformation of the cholesteric layers is more energetically favorable than changing the layer spacing \cite{blanc_helfrich-hurault_2023}. For cholesteric CNC suspensions with a narrow range of pitch values, the undulating layer deformations from the Helfrich-Hurault instability template FCDs. The film also maintains relatively high reflectivity, confirming that the cholesteric layers within each FCD remain locally ordered rather than fully disrupted. However, the angle-resolved spectrum displays a moderate concave shape, reflecting angle dependence. We attribute this angular response to two key factors: (1) planar anchoring at the air and glass interfaces, which enforces a consistent helical axis orientation in significant portions of the film, and (2) the large size of the FCDs (10–100 µm), which allows each domain to behave optically like a macroscopic cholesteric reflector. Thus, while FCD formation introduces curvature into the layers, the underlying cholesteric order is preserved well enough to produce narrowband, specular-like reflection, revealing that defects in CNC films do not necessarily disrupt photonic coherence, but instead modify angular response in a size- and geometry-dependent manner.

The degree of diffuse scattering (and reduction in the angle dependence of the reflected light) of a CNC film is closely tied to its long-range structural disorder. Our spectral analysis demonstrates that the CNC film's photonic response can be tuned through precise adjustments of salt and sonication, which together control suspension viscosity during evaporation. Minimizing viscosity favors cholesteric domain fusion, yielding long-range order and specular, angle-dependent color. Increasing viscosity suppresses tactoid coalescence, promoting structural disorder, which yields diffuse scattering, and more angle-independent coloration. In particular, within films with controlled disorder that have FCDs, curved cholesteric layers redirect reflected light over slightly wider angles compared to defect-free films without increasing the spectral width of the reflected light.

\clearpage

\section{Conclusion}
\indent \indent In this work, we systematically investigated how salt concentration and sonication dose govern the self-assembly, morphology, and photonic response of CNC films. By evaporating CNC suspensions across 24 parameter combinations, we uncovered several key trends. First, tip sonication is most effective at red-shifting the cholesteric pitch when used at low salt concentrations. Second, within a narrow parameter space (around $u_s = 120$ J/mL and [\ce{NaCl}] = 150 mmol/kg), FCDs were produced, and the resulting films reflected submicron wavelengths ($\lambda_{\text{max}} = 737$ nm). Finally, timelapse imaging and viscosity measurements revealed that low-viscosity suspensions produce highly ordered, specular films, underscoring the critical but often overlooked role of viscosity and solvent flow in CNC self-assembly.

These insights demonstrate that CNC film photonic properties can be rationally controlled by tuning salt and sonication, which are simple, accessible processing parameters that modulate viscosity and flow. By manipulating these parameters, we achieved fine control over domain size, pitch, and optical behavior. 

Importantly, we identified a previously unreported regime in which FCDs can be intentionally produced in CNC films. We showed that FCDs emerge only when the viscosity is low enough to produce convection cells during cholesteric nucleation and when kinetic arrest is sufficiently delayed to allow cholesteric layers to remain deformable during late-stage evaporation. These two conditions are met at intermediate salt concentration and moderate sonication, where viscosity balances flow-driven organization with continued layer flexibility. In this regime, evaporative stresses buckle the unarrested cholesteric layers, generating large (50–100 µm) FCDs that preserve local helicoidal order and maintain narrow spectral width. Thus, FCD formation is a controllable structural motif arising from the interplay of viscosity, flow, and kinetic arrest timing.

Achieving both angle-independent coloration and narrow spectral width remains a fundamental challenge. Our findings suggest that tuning viscosity offers a pathway to reduce iridescence without compromising spectral purity. Promising strategies include reducing the size of FCDs and forming FCDs while eliminating planar anchoring from the petri dish, which enforces unwanted alignment and enhances specularity \cite{saha_photonic_2018}. One potential strategy for promoting perpendicular anchoring of CNCs on a substrate could be by leveraging entropic interactions of CNC particles with a geometrically patterned substrate \cite{jull_curvature-directed_2024, campos-villalobos_shaping_2025}. Exploring alternative confinements may also offer new opportunities. For instance, Parker \latin{et al.} \cite{parker_hierarchical_2016, parker_cellulose_2022} produced non-iridescent CNC microparticles in a substrate-free system, though the roles of salt and sonication were not investigated.

Based on these observations, future work should explore how ionic strength, particle morphology, and confinement geometry interact to balance structural uniformity with angular color response. In particular, overcoming planar anchoring may increase control over the size and density of FCDs, offering a path towards minimizing iridescence. This approach may enable the design of CNC-based photonic materials with tailored appearance, ranging from highly specular to fully angle-independent.

\clearpage

\section{Methods}
\subsection{CNC synthesis}
Cellulose nanocrystals were obtained from the acid hydrolysis of Whatman no.1 cellulose filter paper (35 g). The cellulose source was shredded with a spice grinder (Waring WSG60K). We prepared 490 mL of 64 wt.-\% sulfuric acid (Sigma-Aldrich, 95-98\%) and heated it to 66 $^\circ$C. The filter paper was added to the acid solution, stirred vigorously with a mechanical stirrer for 30 minutes during hydrolysis, and the reaction was quenched with five liters of deionized (DI) water. The resulting suspension was allowed to sediment overnight, and the clear supernatant was decanted. The sedimented material was centrifuged in 20-minute cycles at 15000 RCF until it was all in pellet form. This was followed by additional centrifugation cycles, consisting in the addition of DI water until the CNCs in the supernatant were suspended. Only the supernatant was collected and the pellet was discarded. This suspension was dialyzed over two weeks in DI water (replenished daily) with MWCO 12–14 kDa dialysis membranes (Spectrum™ Spectra/Por™ 4). This procedure yielded a 1.1 wt.-\% CNC suspension with a total volume of around 900 mL.

\subsection{Preparation of CNC suspensions} 
The stock CNC suspension was divided and diluted into six separate suspensions (25 mL, [CNC] $=$ 3.5 wt.\%) in 50 mL centrifuge tubes (Corning™ Falcon™). These suspensions were tip sonicated using a Hielscher UP200St ultrasonic processor with a 7 mm titanium tip for various time intervals (corresponding to 0 to 1440 J per mL of CNC suspension) under the following conditions: 30 W, 70:30 ON:OFF cycles, with the tip submerged halfway into the sample. The suspensions were immersed in an ice bath during sonication to prevent sample heating and potential desulfation. For sonication times longer than 5 minutes, 10 minute pauses every 5 minutes were taken to replenish the ice bath and cool the sample. After sonication, each of the six suspensions was divided and diluted (with DI water and a 0.2 M aqueous \ce{NaCl} solution) into four separate suspensions ([CNC] $=$ 3.0 wt.\%) with different \ce{NaCl} concentrations (0 to 450 mmol of \ce{NaCl} per kilogram of CNC) making a total of 24 CNC samples. These CNC suspensions (or films) that underwent a combination of $u_s$ and/or [\ce{NaCl}] will be noted as [``x" J/mL, ``y" mmol/kg] or [``x", ``y"]. For example, [120, 150] or [120 J/mL, 150 mmol/kg] refers to a CNC film with $u_s$ $=$ 120 J/mL and [\ce{NaCl}] $=$ 150 mmol/kg.

\subsection{CNC film formation}
The films presented in this work were evaporated at an initial CNC concentration of 3 wt.\%, in the isotropic phase. This is necessary to have every suspension start the evaporation process equally in the isotropic phase. Each of the 24 CNC suspensions was dropcast (3 mL, 3 wt.\%) into a glass petri dish (3.5 cm in diameter, DURAN® Steriplan®). All CNC suspensions were evaporated simultaneously at around 50-60\% RH and 20 $^\circ$C. The evaporation process usually took 48 to 60 hours to complete.

\subsection{Polarized optical microscopy}
The center of the CNC films (within a 1 cm diameter circle) was imaged in transmission between crossed polarizers on a Nikon Ti-E inverted microscope equipped with a T-P2 DIC Polarizer module and an analyzer. The images and videos were captured with a Nikon DS-Ri2 CMOS camera. The timelapse videos of evaporating CNC suspensions were recorded at 1 frame per minute. The videos were captured under software-side light auto-exposure, varying from 1 second at evaporation start to around 30 ms on evaporation completion. This was done to assure adequate light exposure when recording any birefringence, at the cost of over-exposing frames with no birefringence. 

\subsection{Optical microscopy}
The center of the colored films was imaged in reflection, with a left circular polarizer in the light path, on a Leica DM IL LED inverted microscope. The photographs were taken with a mounted Nikon Z6 digital camera.

\subsection{UV-Vis transmission spectrometry}
UV-Vis transmission spectra for the center of the films were obtained using a LAMBDA 365 UV/Vis spectrophotometer. For all spectra, the acquisition rate was 960 nm/min, in a wavelength range of 400 to 1100 nm. For each film, a rectangular region of about 1 cm by 0.5 cm was analyzed in its center. 

\subsection{Scanning electron microscopy (SEM)}
The morphological properties of CNC films were observed using a Thermo Scientific Phenom™ XL G2 Desktop scanning electron microscope at 15 kV and 60 Pa, with a working distance of 6 to 7 mm. To observe the inner structure of the CNC films, we submerged them in liquid nitrogen for one minute and then fractured them with tweezers to expose the cross section.

\subsection{Angle-resolved spectroscopy}
Angle-resolved reflectivity spectra were taken with a Nikon halogen lamp and a 60$\times$ air objective (Nikon ELWD, CFI Plan Fluor, NA=0.7), at a wavelength interval of 450 to 800 nm. For each film, small and large circular regions (20 {\textmu}m and 200 {\textmu}m in diameter, respectively) were analyzed by varying the opening of the aperture. 

\subsection{Viscosity measurement}
The rheological measurements were performed with a rotational rheometer (MCR300, Anton Paar), using cone-plate geometry. The cone has a diameter of 50 mm, a cone angle of 1.001$^\circ$, and a truncation of 0.051 mm. For all measurements, the temperature was set at 25 $^\circ$C and the applied sample volume was 1 mL (before trimming) to ensure full coverage. Each sample was analyzed in a shear rate interval of 0.1 to 150 s$^{-1}$, taking 50 data points over 5 minutes. Data points were chosen only if the torque value was above the recommended 0.1 mN.m. Care was taken to ensure that no significant solvent evaporation occurred.

\clearpage
\begin{acknowledgement}

The authors thank Dominique Thies-Weesie for assistance with viscometry measurements, Chris Schneijdenberg for assistance with SEM imaging, and Dave van den Heuvel and Relinde van Dijk-Moes for helpful discussions and technical assistance. L.T. acknowledges useful discussion with Rae Robertson-Anderson. D.V.S. and I.R.V. acknowledge financial support from the Department of Physics, Utrecht University. L.T. acknowledges support from the European Commission (Horizon-MSCA, Grant No. 892354) and the Dutch Research Council NWO ENW Veni grant (Project No. VI.Veni.212.028). D.V.S. and L.T. acknowledge support from the Starting PI Fund for Electron Microscopy Access from Utrecht University’s Electron Microscopy Center. F.T.R. acknowledges support from the Dutch Research Council
NWO (OCENW.KLEIN.008 and Vi.Vidi.203.031).

\end{acknowledgement}

\clearpage
\begin{suppinfo}

\setcounter{figure}{0}
\makeatletter\renewcommand{\thefigure}{S\arabic{figure}}\makeatother

\section{Supporting Information}

\subsection{Complete grid with LCP reflection images}
\indent \indent Reflection images (through a left circular polarizer) were captured for each CNC film. This showcases all 24 CNC films, including the ones that do not exhibit any visible structural color.
\begin{figure}[H]
\begin{center}
  \includegraphics[width=1\linewidth]{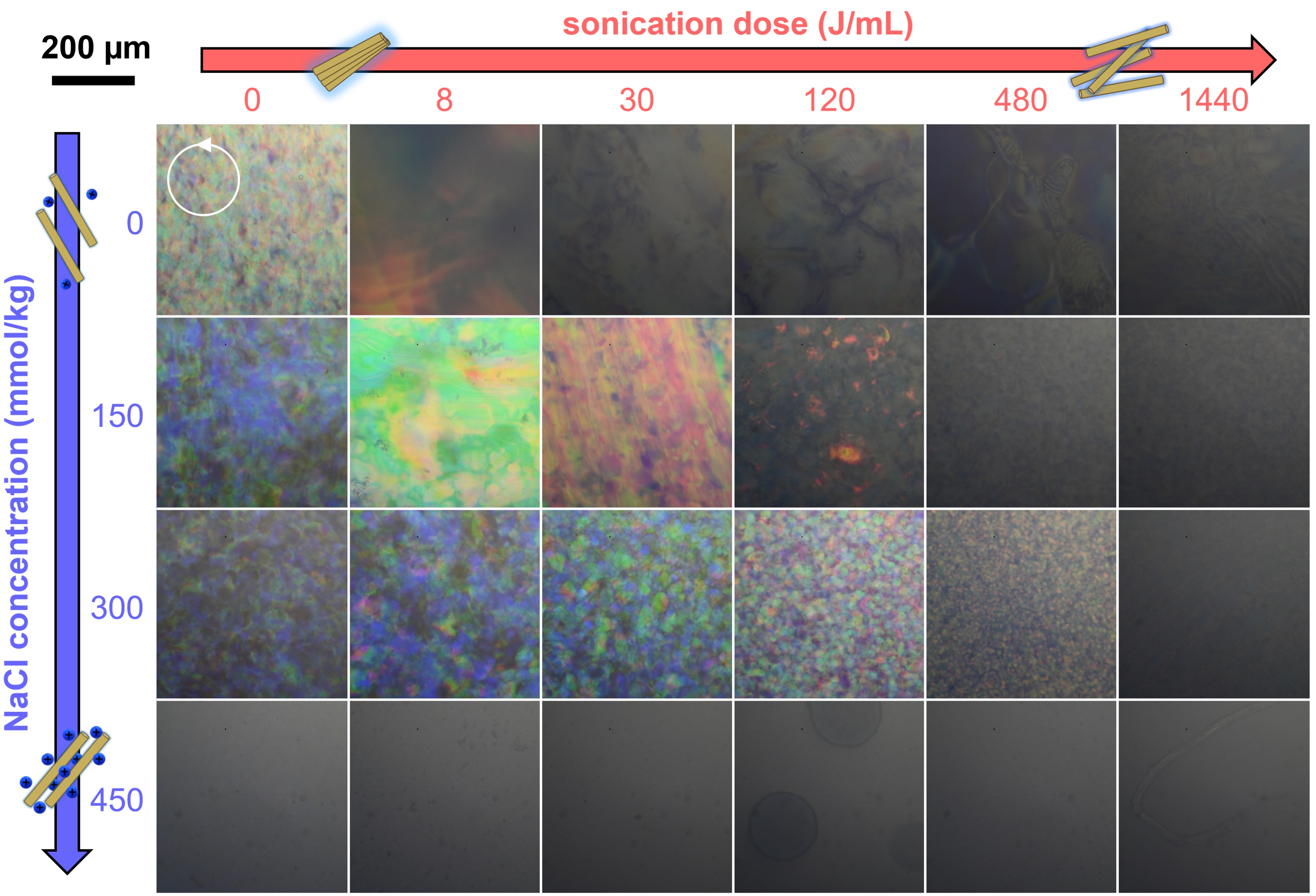}
  \end{center}
  \caption{An array of microscope images taken in reflection mode, captured through a left circular polarizer. All pictures were taken within a 5 mm region around the center of the film.}
  \label{si_lcp}
\end{figure}

\clearpage 
\subsection{Photographs of CNC films}
\indent \indent Macroscopic photographs of a selection of films were captured with the camera of a Samsung Galaxy S21 FE. The films were illuminated with a white ring light behind the camera and photographed at a 12$^\circ$ angle with the substrate. In SI Figure \ref{si_photo}, we show most of the CNC films that display visible structural color at a macroscopic scale.
\begin{figure}[H]
\begin{center}
  \includegraphics[width=1\linewidth]{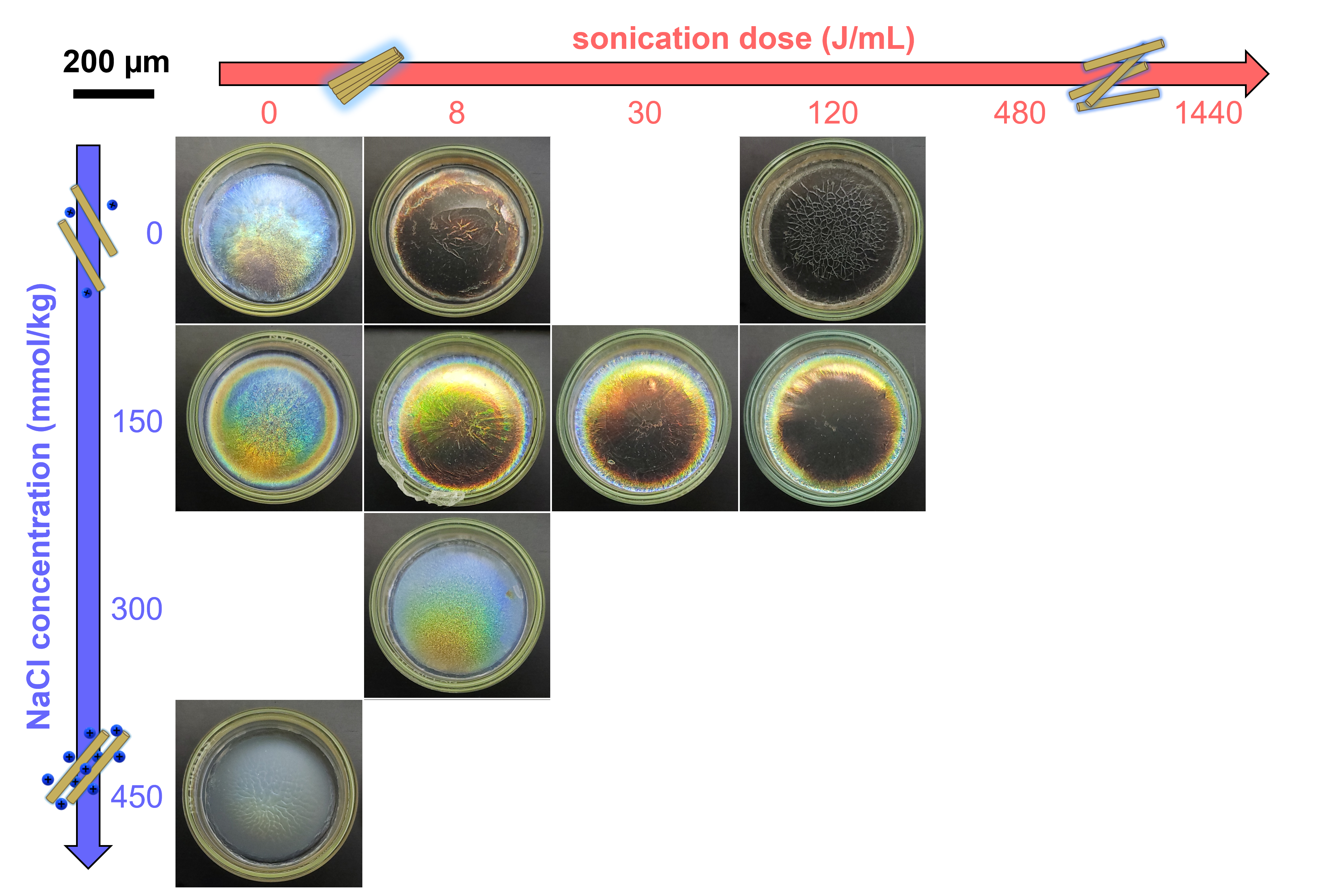}
  \end{center}
  \caption{An array of photographs of CNC films presented in this work. For scale, the diameter of all petri dishes is 3.5 cm.}
  \label{si_photo}
\end{figure}
\noindent The majority of the films develop a ``coffee ring effect". Due to the significant capillary flow during evaporation, the suspension becomes more concentrated at the edge. This then translates to a higher thickness and a blue shift towards its perimeter. This effect is mostly reported in evaporated sessile droplets \cite{girard_effect_2008, gencer_influence_2017, shao_marangoni_2021}, but is also present in freestanding films \cite{dumanli_digital_2014}.

\clearpage
\subsection{Estimation of Debye length for different salt concentrations}
\indent \indent The ionic strength of a given solvent can be used to estimate its Debye length. Considering the solvent is water at 20 $^\circ$C, the Debye length $\kappa^{-1}$ is given by the equation:

\begin{equation}
    \kappa^{-1} = (\frac{\epsilon_0 \epsilon_r k_B T}{2 c q_e^2})^{\frac{1}{2}},
    \label{eq_debyelength}
\end{equation}

\noindent where $\epsilon_0$ is the vacuum permittivity, $\epsilon_r$ is the dielectric permittivity of the solvent ($\epsilon_r \approx 2.38$ for toluene), $k_B$ is the Boltzmann constant, $T$ is the solution temperature, $q_e$ is the charge of an electron, and $c$ is the ionic concentration (in mol/L), which accounts for both [H+] and [Na+] ions in solution. At the initial CNC concentration of 3 wt.\%, [H+] = 3.2 mM (calculated from pH = 2.5, as measured from the stock CNC suspension). and [Na+] = 0 mM, 4.5 mM or 9.0 mM (for [\ce{NaCl}] = 0, 150 or 300 mmol/kg), respectively. In SI Figure \ref{si_debye}, we plot the Debye length against increasing CNC concentrations at different [\ce{NaCl}] values, as to emulate a CNC film evaporation.

\begin{figure}[H]
\begin{center}
  \includegraphics[width=0.5\linewidth]{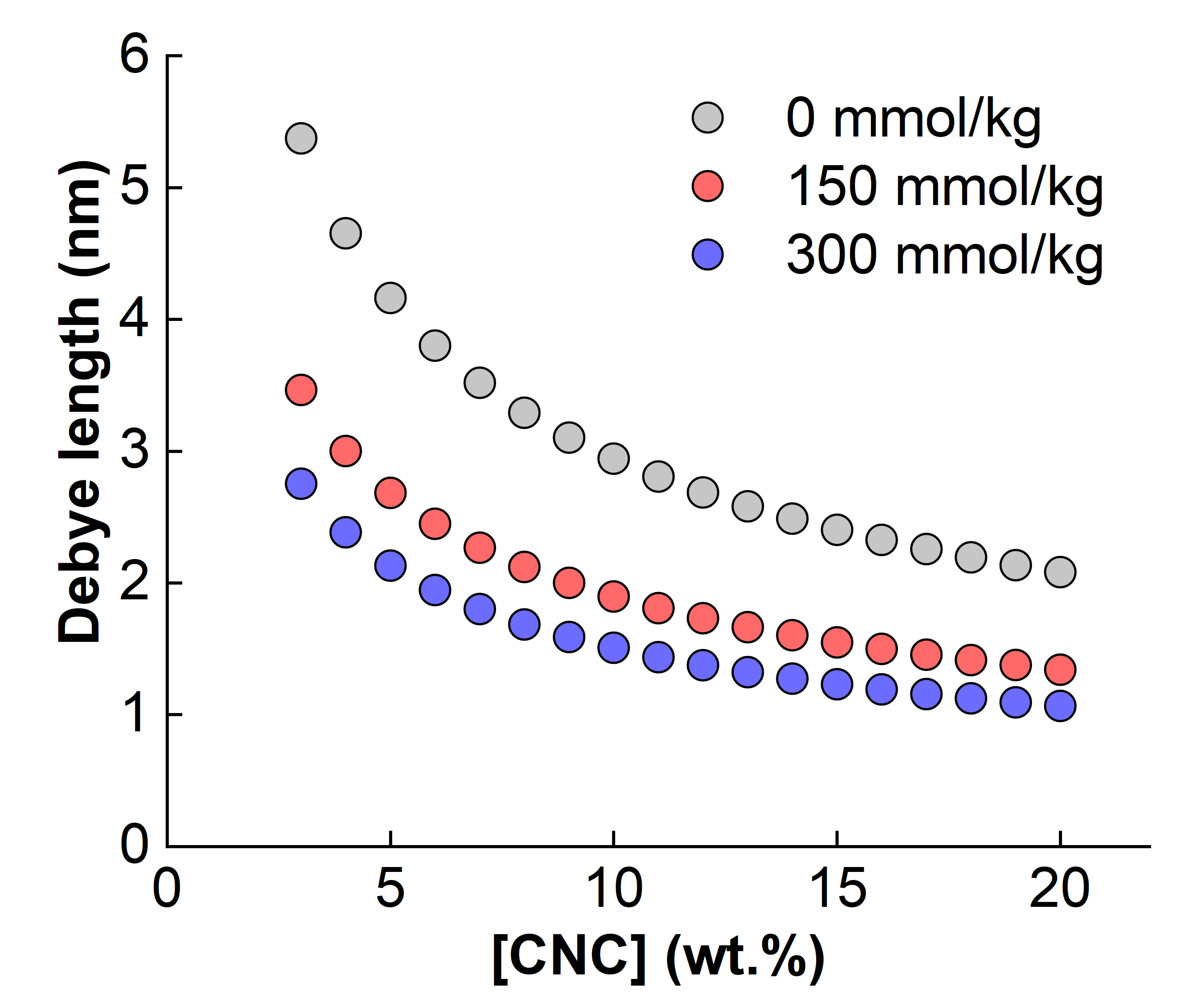}
  \end{center}
  \caption{Debye length (y-axis) for given CNC concentration values from 3 to 20 wt.\% (x-axis), at different [\ce{NaCl}] values.}
  \label{si_debye}
\end{figure}

\clearpage
\subsection{UV-Vis spectra spectral width}
\indent \indent To quantify the color homogeneity of the CNC films shown in Figure \ref{ch2:fig1}, their Full Width at Half Maximum values (FWHM, also referred to as spectral width) were estimated.

\begin{table}[!ht]
    \centering
    \begin{tabular}{|c|c|c|c|c|c|c|}
    \hline    &\textbf{0 J/mL}& \textbf{8 J/mL} & \textbf{30 J/mL} & \textbf{120 J/mL} & \textbf{480 J/mL} & \textbf{1440 J/mL}\\ 
    \hline  \textbf{0 mmol/kg}  &151 & 382 & 672 & N/A& N/A & N/A\\ 
    \hline
 \textbf{150 mmol/kg}& 206& 161& 100& 181& 240&N/A\\\hline
 \textbf{300 mmol/kg}& N/A& 134& 87& 127& 204&N/A\\\hline
    \end{tabular}
\caption{Array of estimated spectral width values (in nanometers) from UV-Vis reflectance spectra. For spectra marked with ``N/A", their peak wavelength was outside of the scanned wavelength range of 400 to 1100 nm, rendering it impossible to estimate their spectral width reliably.}
\label{si_fwhm}
\end{table}

\noindent For [\ce{NaCl}] $=$ 0 mmol/kg, sonication dose quickly increases FWHM values, indicating a wide range of wavelengths is reflected. Then, for the remaining films at [\ce{NaCl}] $=$ 150 mmol/kg and [\ce{NaCl}] $=$ 300 mmol/kg, there appears to be a minimum at $u_s$ $=$ 30 J/mL. Though spectral width is correlated with disorder in a film's cholesteric structure: the wider the range of reflected wavelengths, the larger the variance in cholesteric domain size, angle, and pitch value.

\clearpage
\subsection{Effect of suspension filtration on CNC film assembly}
\indent \indent To observe the effect of large CNC aggregates on a film's homogeneity, while avoiding the use of ultrasonication, we evaporated two films from the same [0 J/mL, 150 mmol/kg] solution. Before dropcasting and evaporation, one of the suspensions was filtered with a Nalgene™ syringe filter (cellulose acetate, pore size 0.8 {\textmu}m). The filtration step is meant to eliminate CNC particles with sizes above 800 nm, which may negatively impact the homogeneity of the cholesteric structure. POM images of the unfiltered and filtered film are shown in SI Figure \ref{si_filter}A, B.

\begin{figure}[H]
\begin{center}
  \includegraphics[width=0.75\linewidth]{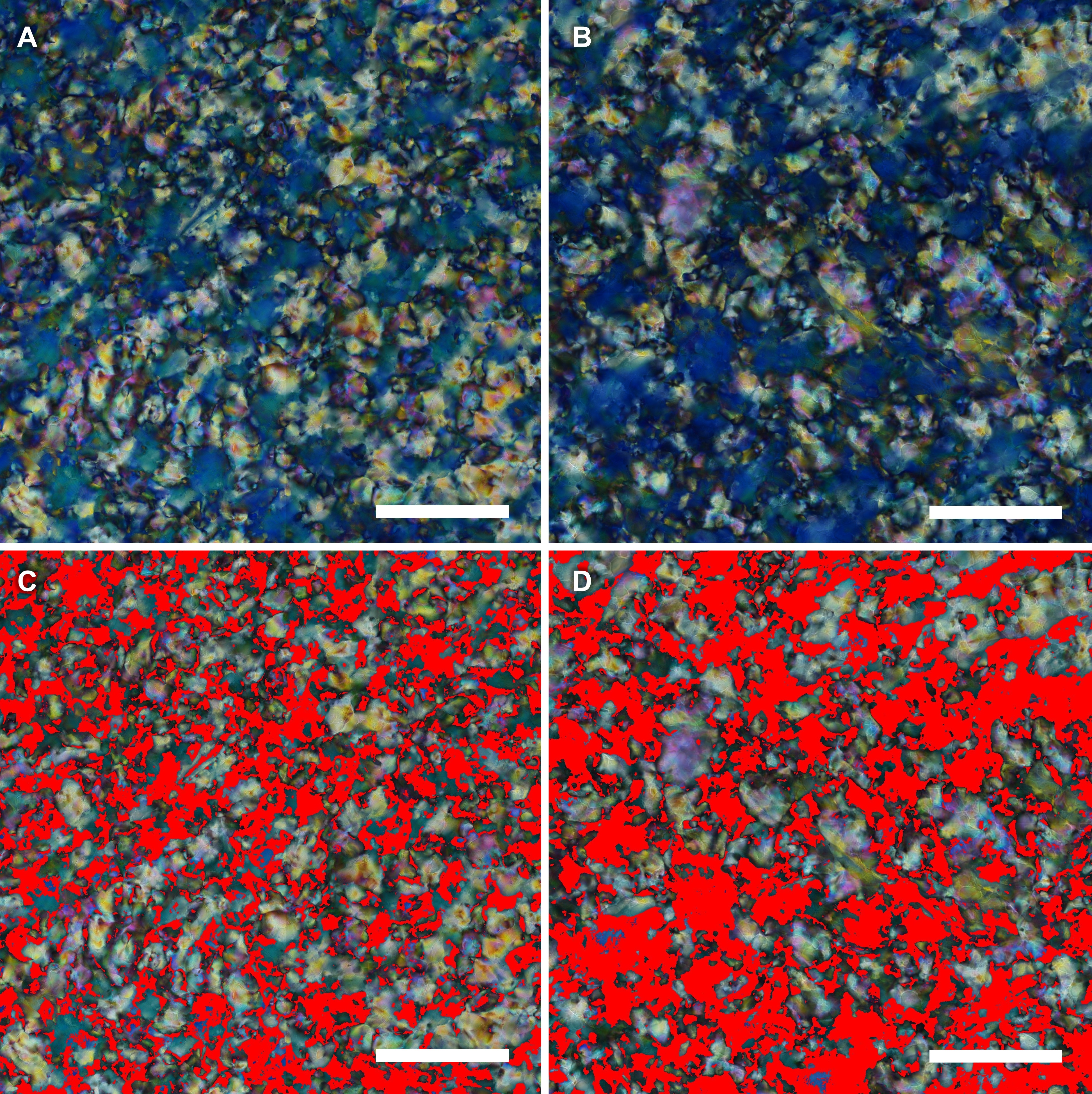}
  \end{center}
  \caption{Stitched micrographs (2$\times$2 regions) of (A) unfiltered and (B) filtered [0 J/mL, 150 mmol/kg] CNC films. (C) and (D) are equivalent to (A) and (B), but all blue hue pixels were masked in red with ImageJ. Scale bars: 500 {\textmu}m.}
  \label{si_filter}
\end{figure}

By visually comparing the unfiltered and filtered films (SI Figure \ref{si_filter}A, B), the filtered film appears to have more blue uniform domains. This is confirmed by using ImageJ to apply a color threshold to each image (SI Figure \ref{si_filter}C, D). We set the following HSB levels: hue from 140 to 200, saturation from 140 to 255 and brightness from 0 to 140. These parameters highlight the distinctly uniform, blue regions. By counting all of the ``blue" pixels, the unfiltered film's image was calculated to have 31\% of its area covered in the selected tones of blue, while the filtered film has 42\% of its area in blue. So, quantitatively, the filtered film shows more color homogeneity and structural uniformity due to its filtered-out CNC aggregates.

\clearpage
\subsection{[120 J/mL, 150 mmol/kg] film: slow evaporation}
\indent \indent A film made from the [120 J/mL, 150 mmol/kg] suspension was slowly evaporated while under POM observation. Its composition and evaporation method were identical to all other films, except it was loosely covered with a glass lid for the duration of the evaporation in order to slow it down. This extended the evaporation time from around 2 days (uncovered) to 11 days. In SI Figure \ref{si_slow}, we show macro- and microscopic surface images of the produced film, its reflectance spectrum, and its exposed cross-section taken with SEM.
\begin{figure}[H]
\begin{center}
  \includegraphics[width=0.8\linewidth]{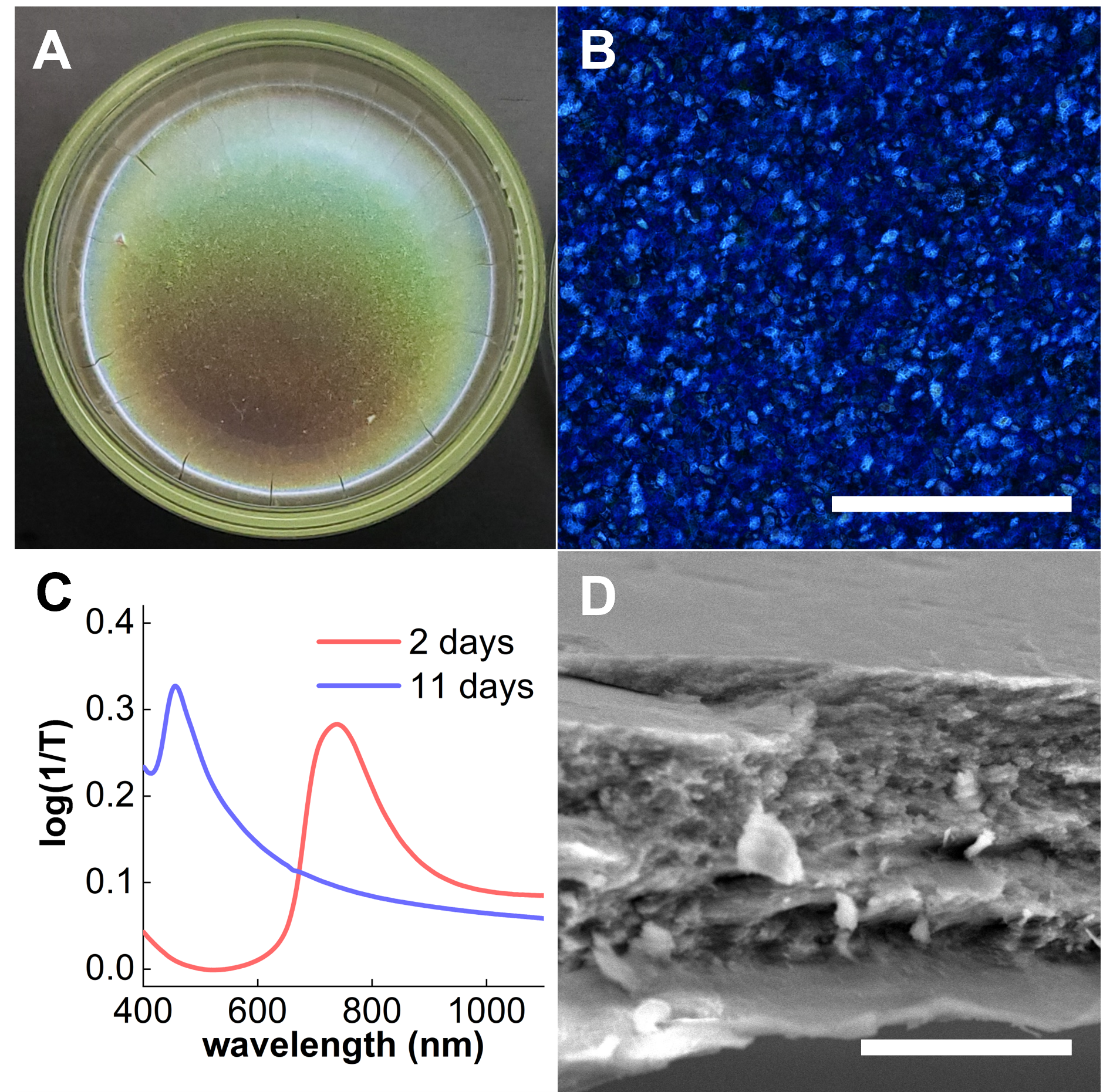}
  \end{center}
  \caption{(A) Photograph of the entire dish-cast film. The petri dish diameter is 3.5 cm. (B) POM image of the film. (C) Reflectance spectra comparing films evaporated from a [120 J/mL, 150 mmol/kg] suspension with differing evaporation times (2 and 11 days). (D) Cross-section image of the film, taken with SEM. Scale bars: b) 500 {\textmu}m, d) 20 {\textmu}m.}
  \label{si_slow}
\end{figure}

\clearpage
\subsection{CNC size distribution}
\indent \indent Atomic Force Microscopy (AFM) imaging was utilized to verify the effect of ultrasonication on CNC length. In preparation for AFM imaging, we used a freshly cleaved mica sheet (Ted Pella, V1 Mica, 25 $\times$ 25mm) as a substrate. Its surface was positively charged by dropcasting 20 {\textmu}L of poly-L-lysine (PLL, 0.01 wt.\%) and letting it react for 45 seconds. Afterward, the surface was thoroughly washed with Millipore water and air-dried with nitrogen. Next, 200 {\textmu}L of 0.004 wt.-\% CNC suspension were dropcast onto the substrate, allowed to settle for 45 seconds, then washed with Millipore water and air-dried. This process was done for an unsonicated CNC suspension and replicated for four other CNC suspensions with different applied sonication doses (8, 30, 120 and 1440 J/mL). \\

The AFM imaging was performed with a Nanowizard ULTRA Speed 2 (Bruker), with Bruker FASTSCAN-A tips (scanning frequency 1.4 MHz, k $=$ 17 N/m) in tapping mode. From the obtained AFM images, for each sonication step, the lengths of 600 particles were measured using the software Gwyddion (version 2.62, http://gwyddion.net/). SI Figure \ref{si_afm} shows the measured particle length distribution in a box plot graph.
\begin{figure}[H]
\begin{center}
  \includegraphics[width=0.5\linewidth]{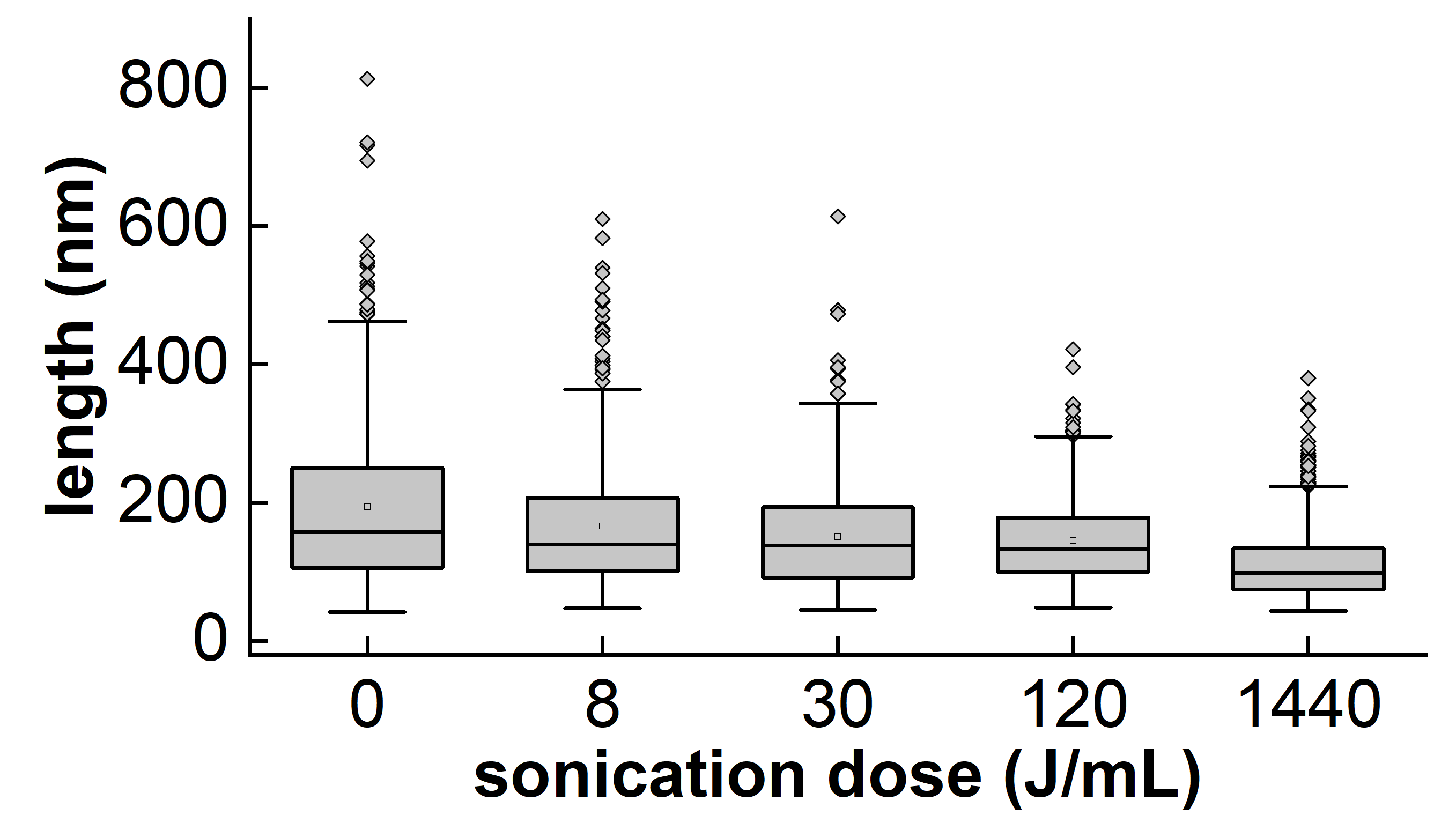}
  \end{center}
  \caption{A box plot graph of CNC particle length values, illustrating the decrease in average particle length with sonication dose.}
  \label{si_afm}
\end{figure}
The CNC hydrodynamic size at different sonication steps was measured through Dynamic Light Scattering (DLS). Measurements were performed at a CNC concentration of 0.15 wt.\% (in a DTS1070 Capillary Zeta Cell). For each sample, measurements were performed in three cycles of 20 runs each, using a Malvern Nano Zetasizer.

\begin{figure}[H]
\begin{center}
  \includegraphics[width=0.5\linewidth]{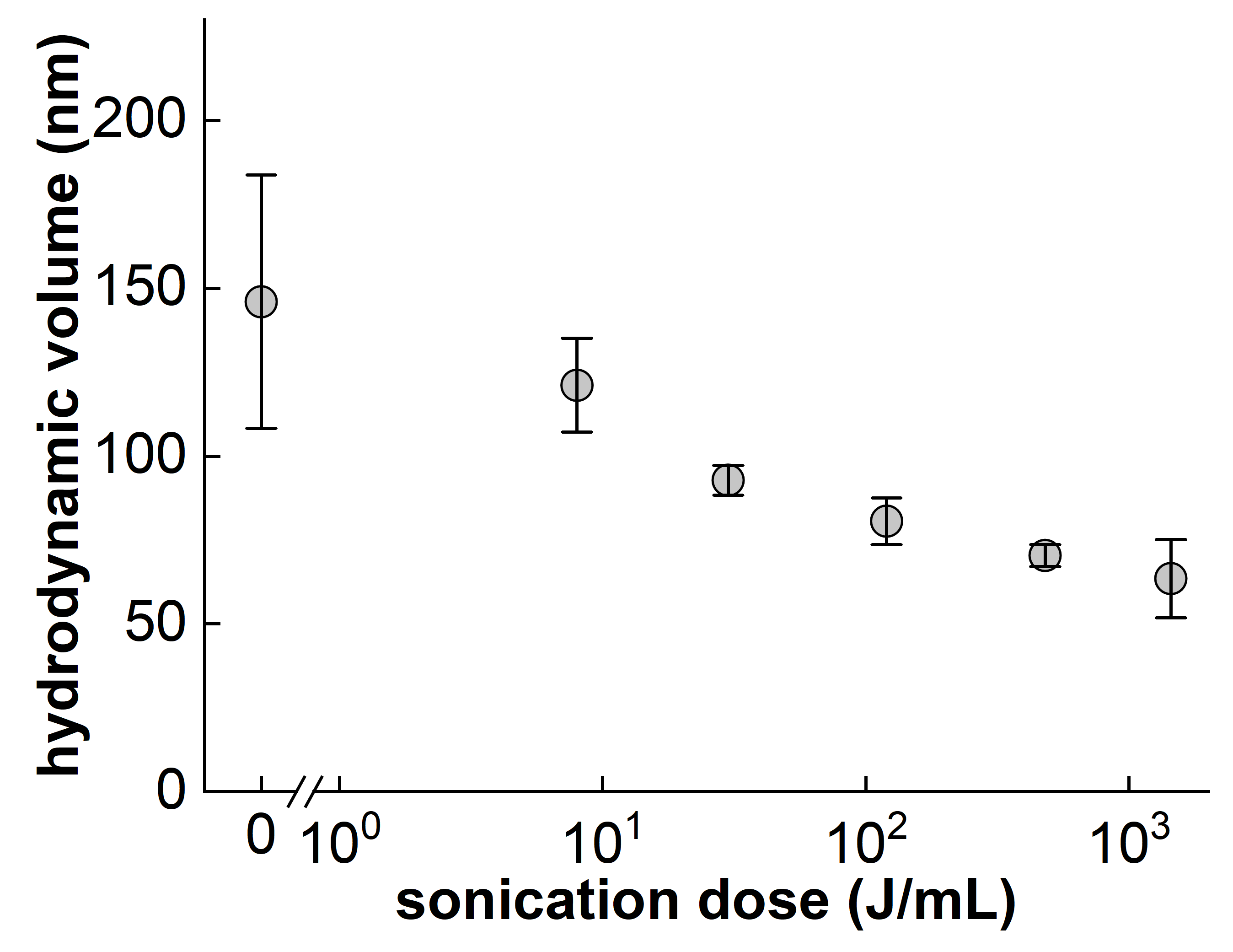}
  \end{center}
  \caption{Hydrodynamic volume as measured by DLS, illustrating a decrease in average particle size with sonication dose.}
  \label{si_dls}
\end{figure}
Both methods show a clear decrease in CNC particle size with sonication dose, due to the fragmentation of CNC bundles and particles.

\clearpage
\subsection{Calculation of isotropic and cholesteric phase densities}
\indent \indent To estimate the difference in density ($\Delta\rho$) between the cholesteric and the isotropic phase, 4 mL of a [30 J/mL, 150 mmol/kg] suspension (12 wt.\%) was left to equilibrate in a glass vial over two weeks. A cholesteric-isotropic biphase formed, at a  50:50 volume fraction. 100 {\textmu}L from each phase was carefully pipetted and weighed. The top isotropic phase weighed 103.2 mg ($\rho$ = 1.032 $\times$ 10$^6$ g/m$^3$) and the bottom cholesteric phase weighed 106.5 mg ($\rho$ = 1.065 $\times$ 10$^6$ g/m$^3$ ). Thus, $\Delta\rho$ was calculated to be 3.3 $\times$ 10$^4$ g/m$^3$.

\clearpage
\subsection{Conductivity measurements}
\indent \indent Tip sonication of CNC samples has been shown to increase the electrical conductivity of the suspension, which is proportional to its ionic content \cite{beck_controlling_2011, parton_chiral_2022}. The increase in ionic conductivity occurs through the release of water-bound ions on CNCs into the suspension, leading also to a decrease in pH. This consequence makes it difficult to decouple the effect of sonication from the effect of added ionic content.

We measured the conductivity of select CNC samples to compare the magnitude of the increase in conductivity that stemmed from sonication and salt concentration. Conductivity measurements were performed on 7 mL suspensions, under ambient conditions, using a Eutech CON 450 conductivity meter. All suspensions had a CNC concentration of 3.5 wt.\%. These values are presented in SI Table \ref{si_conductivity}.

\medskip 
\begin{table}[!ht]
    \centering
    \begin{tabular}{|c|c|}
    \hline  \textbf{Sample}& \textbf{Conductivity (mS/cm)}\\ 
    \hline  [0 J/mL, 0 mmol/kg]&1.20\\ 
    \hline  [8 J/mL, 0 mmol/kg]&1.22\\ 
    \hline  [1440 J/mL, 0 mmol/kg]&1.41\\
    \hline  [0 J/mL, 150 mmol/kg]& 3.28\\
    \hline  [8 J/mL, 150 mmol/kg]& 3.20\\
    \hline  [30 J/mL, 150 mmol/kg]& 3.25\\
    \hline  [120 J/mL, 150 mmol/kg]& 3.31\\
    \hline  [8 J/mL, 300 mmol/kg]& 4.26\\
    \hline
    \end{tabular}
\caption{List of samples for which the electrical conductivity was measured, along with their respective conductivity values.}
\label{si_conductivity}
\end{table} 

With no added \ce{NaCl}, the conductivity remains low, at a baseline of around 1.2 mS. As expected, this value is slightly raised with increased sonication (up to 1.41 mS/cm for $u_s$ = 1440 J/mL). Adding salt ([\ce{NaCl}] $=$ 150 mmol/kg) raises the measured conductivity much more effectively than through sonication: for the lowest step of added \ce{NaCl}, the conductivity increases from approximately 1.2 to 3.2 mS/cm.  Additionally, at this [\ce{NaCl}], applying tip-sonication does not modify the solution's conductivity in any discernible way. This shows that the effect of one step of added salt is much greater than any step of applied sonication (within the scope of this work) in regards to increasing ionic content in the CNC suspension.
\clearpage

\subsection{Viscosity fits at [CNC] $=$ 8 wt.\% }
\begin{figure}
\begin{center}
  \includegraphics[width=0.6\linewidth]{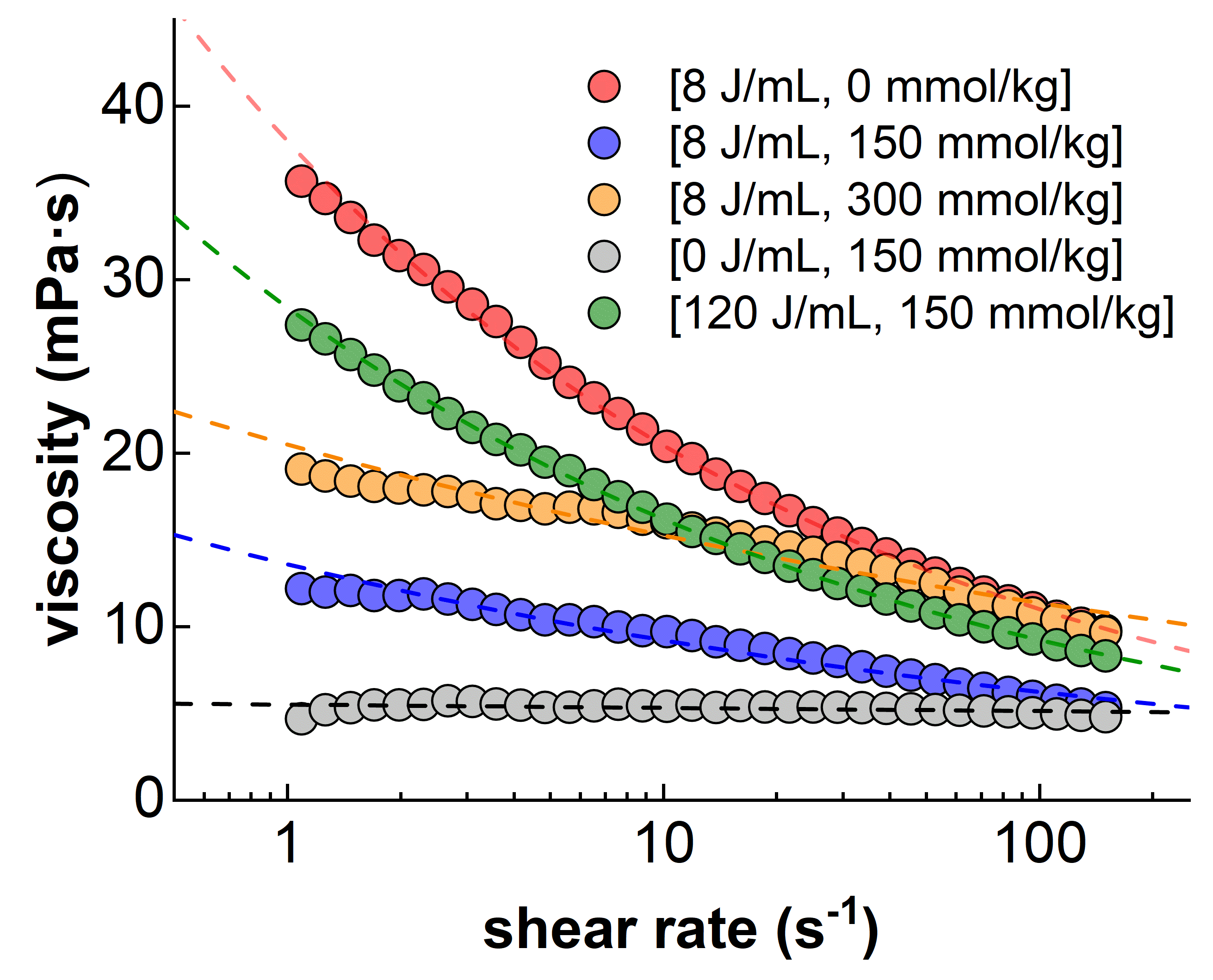}
  \end{center}
  \caption{Viscosity of 8 wt.\% CNC suspensions as a function of shear rate, fitted to the power-law model (Eq. \ref{eq:powerlaw}). Experimental data (circles) and model fits (dashed lines) are shown for all salt and sonication conditions. The fits were used to extract the flow behavior index ($n$) and flow consistency index ($K$) values summarized in Table \ref{ch2:viscosity_table}.}
  \label{SI_viscosityfits}
\end{figure}

\clearpage
\subsection{All angle-resolved spectroscopy maps}

\begin{figure}
\begin{center}
  \includegraphics[width=0.9\linewidth]{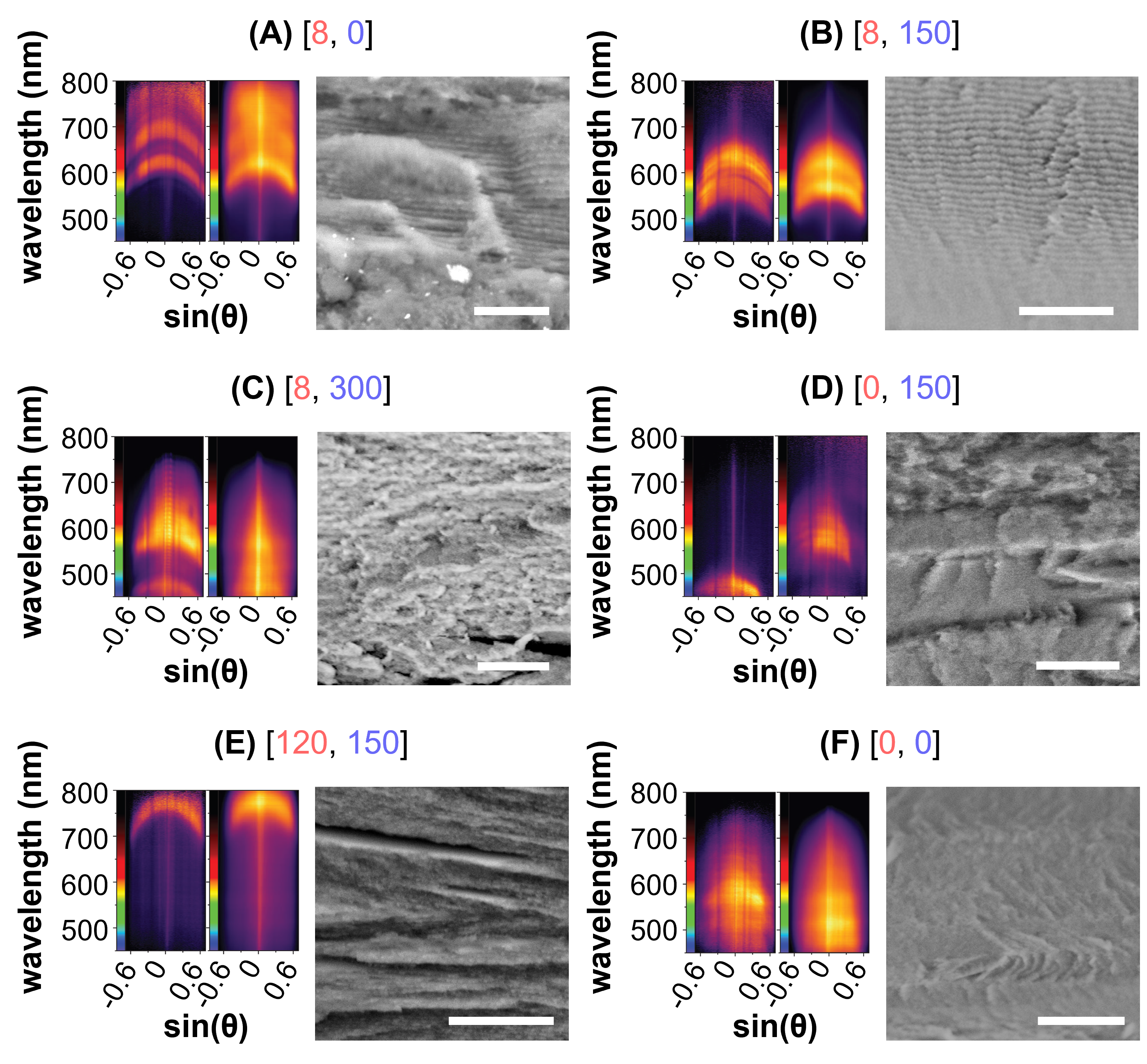}
  \end{center}
  \caption{For each of six films (A-F), angle-resolved reflectance spectra were obtained for a small region (20 {\textmu}m diameter, left) and a large region (200 {\textmu}m diameter, right). The rightmost image for each film is its corresponding exposed cross-section taken with SEM. (A) [8 J/mL, 0 mmol/kg], (B) [8 J/mL, 150 mmol/kg], (C) [8 J/mL, 300 mmol/kg], (D) [0 J/mL, 150 mmol/kg], (E) [120 J/mL, 150 mmol/kg], (F) [0 J/mL, 0 mmol/kg]. SEM image scale bars: 10 {\textmu}m.}
  \label{SI_allangleresolved}
\end{figure}

\clearpage 
\subsection{Timelapse videos}

\indent Timelapse videos of different evaporating CNC suspensions were recorded with polarized optical microscopy. All of the videos provided are listed below.

\begin{itemize}
    \item SI Video 1 (1m08s) - Evaporation timelapse of sample [120 J/mL, 150 mmol/kg].
    \item SI Video 2 (1m39s) - Evaporation timelapse of sample [120 J/mL, 150 mmol/kg] at a slowed evaporation rate. The video starts at the 188-hour mark (still in the isotropic regime).
    \item SI Video 3 (1m27s) - Evaporation timelapse of sample [0 J/mL, 150 mmol/kg].
    \item SI Video 4 (1m07s) - Evaporation timelapse of sample [8 J/mL, 150 mmol/kg].
    \item SI Video 5 (1m29s) - Evaporation timelapse of sample [30 J/mL, 150 mmol/kg].
    \item SI Video 6 (0m56s) - Evaporation timelapse of sample [0 J/mL, 0 mmol/kg].
    \item SI Video 7 (0m43s) - Evaporation timelapse of sample [1440 J/mL, 0 mmol/kg].
    \item SI Video 8 (1m13s) - Evaporation timelapse of sample [0 J/mL, 450 mmol/kg].
\end{itemize}

\end{suppinfo}

\clearpage
\bibliography{references}

\end{document}